\documentclass[aip,jcp,preprint]{revtex4-1}
\usepackage{graphicx}
\graphicspath{{./Figures/}}
\usepackage{dcolumn}
\usepackage{bm}
\usepackage[mathlines]{lineno}
\usepackage[utf8]{inputenc}
\usepackage[T1]{fontenc}
\usepackage{mathptmx}
\usepackage{fullpage}
\usepackage{amssymb}
\usepackage{amsmath}
\usepackage{mathrsfs}
\usepackage{multirow}
\usepackage{float}
\usepackage{subcaption}
\usepackage[font=small,labelfont=bf,
   justification=justified,
   format=plain]{caption}

\newcommand{\bra}[1]{\langle #1 |}
\newcommand{\ket}[1]{| #1 \rangle}

\newcommand{\bk}[2]{\bra{#1} #2 \rangle}

\newcommand{\dd}[2]{\frac{d #1}{d #2}}

\newcommand{\no}{\nonumber}

\newcommand{\mco}{\mathcal{O}}
\newcommand{\mcob}[1]{\mathcal{O}(#1)}

\newcommand{\eqr}[1]{Eq.~\eqref{eq:#1}}

\newcommand{\eql}[1]{\label{eq:#1}}
\newcommand{\figr}[1]{Fig.~\ref{fig:#1}}
\newcommand{\figl}[1]{\label{fig:#1}}

\begin{document}
\raggedbottom


\title{Electronic energies from coupled fermionic 'Zombie' states imaginary time evolution}
\author{Oliver A.\ Bramley}
\email{cm14oab@leeds.ac.uk}
\affiliation{School of Chemistry, University of Leeds, Leeds, LS2 9JT, UK}
\author{Timothy J.\ H.\ Hele}
\email{t.hele@ucl.ac.uk} 
\affiliation{Department of Chemistry,University College London, WC1H 0AJ, UK}
\author{Dmitrii V.\ Shalashilin}
\email{d.shalashilin@leeds.ac.uk}
\affiliation{School of Chemistry, University of Leeds, Leeds, LS2 9JT, UK}

\date{\today}
\bibliographystyle{aip}
\begin{abstract}
Zombie States are a recently introduced formalism to describe coupled coherent Fermionic states which address the Fermionic sign problem in a computationally tractable manner. Previously it has been shown that Zombie States with fractional occupations of spin-orbitals obeyed the correct Fermionic creation and annihilation algebra and presented results for real-time evolution \cite{RN106}. In this work we extend and build on this formalism by developing efficient algorithms for evaluating the Hamiltonian and other operators between Zombie States and address their normalization. We also show how imaginary time propagation can be used to find the ground state of a system. We also present a biasing method, for setting up a basis set of random Zombie States, that allow much smaller basis sizes to be used while still accurately describing the electronic structure Hamiltonian and its ground state and describe a technique of wave function "cleaning" which removes the contributions of configurations with the wrong number of electrons, improving the accuracy further. We also show how low-lying excited states can be calculated efficiently using a Gram-Schmidt orthogonalization procedure.The proposed algorithm of imaginary time propagation on a biased random grids of Zombie States may present an alternative to existing Quantum Monte Carlo methods.
\end{abstract}

\maketitle

\section{Introduction} 

The wavefunction of a multi-electron system is usually accurately described by a linear combination of Slater determinants\cite{RN111}. Each Slater determinant represents a specific configuration --- how the available electrons are distributed across spin-orbitals. However, in a system with many electrons and many orbitals the number of possible Slater determinants increases rapidly. Even with sensible approximations and truncation, accurately describing the electronic structure of a system requires a large number of Slater determinants and with that comes computational cost. Introducing an active space of a few most important orbitals and few active electrons is the main method of treating this problem. Active space removes abruptly all configurations outside of it, but even though the contribution of individual configurations outside of the active space can be small, their total contribution can be significant.  A smooth transition between active and virtual orbitals in active space methods can potentially be beneficial. Such a transition can be perhaps achieved with Quantum Monte-Carlo methods.  Diffusion MC\cite{RevModPhys.73.33} and Green's function MC\cite{PhysRev.128.1791} propagate walkers in continuum space which removes the need for a one-electron basis set but they both fall foul of the fermionic sign problem caused by the antisymmetry of the electronic wavefunction. It has been possible to solve this problem using a fixed node approximation\cite{doi:10.1063/1.431514} but this has had limited further work. A more robust approach is to introduce a Monte-Carlo way of selecting the configurations via a random walk in the manifold of Slater Determinants. Full Configuration-Interaction Quantum Monte-Carlo (FCIQMC), developed by Alavi and co-workers, works by using a long-time propagation in imaginary time and random walkers to stochastically describe the full configuration-interaction wavefunction (FCI) \cite{doi:10.1063/1.3193710, doi:10.1063/1.3302277}. The method does not converge to the Bosonic solution \cite{doi:10.1063/1.3193710} and obtains the Fermionic ground state without a fixed node approximation. FCIQMC has been successful at obtaining FCI results for large systems which were not previously possible including the neutral and cationic elements from Li to Mg \cite{doi:10.1063/1.3193710,doi:10.1063/1.3407895,doi:10.1063/1.3302277,doi:10.1063/1.3525712,doi:10.1063/1.3624383}. A similar method is Monte Carlo Configuration Interaction (MCCI), developed by Greer\cite{doi:10.1063/1.469756, TONG2000142}, which originally was applied to the single point energy of water\cite{doi:10.1063/1.469756} and later used to find the dissociation energy of water and HF\cite{doi:10.1063/1.4767052}. Like FCIQMC, MCCI uses a Monte Carlo procedure to build a compact wavefunction containing the important configurations intended to give accuracy close to FCI results. Coe and Paterson \textit{et. al} have continued to develop MCCI and shown that it can generate potential energy curves, approaching levels of chemical accuracy, for a range of small molecules including N$_2$ and CH$_4$\cite{doi:10.1063/1.4767052}. MCCI was then extended to include Natural obitals and second-order perturbation theory which, particularity, at longer bond lengths was shown to improve accuracy and convergence time of calculated energies\cite{doi:10.1063/1.4767436}. MCCI has also been used to simulate multipole moments achieving good accuracy when compared to FCI values\cite{https://doi.org/10.1002/jcc.23211}. MCCI also showed good accuracy when compared to FCIQMC when calculating ionization energies while only requiring a relatively small number of basis functions when compared to the FCI space\cite{https://doi.org/10.1002/jcc.23211}. The method has also been extended to include state-averaging to allow computation of excited states which was shown to be effective for both H$_3$ and LiF\cite{doi:10.1063/1.4824888}.

However, when used as part of a molecular dynamics simulation the computational cost of Electronic structure calculations, especially Monte-Carlo methods, is still significant. Hence, finding methods with acceptable cost of electronic structure calculations, which at the same time would not neglect configurations that normally would be outside of the active space, is of great benefit. In a recent paper\cite{RN106} we introduced the idea of a Zombie States (ZS) which potentially can combine the ideas of randomness and active space. Zombie states have the possibility of an electron in every orbital in a superposition of "dead" and "alive" states \cite{RN106} so that the occupation of an orbital is fractional.  This allows multiple Slater determinant configurations to be described by a single ZS.  However, ZS may not have a well-defined number of electrons, such that there is a nonzero probability of the state containing a wrong number of electrons. But a small superposition of ZSs might be able to describe a wave function with the required number of electrons accurately, and the hope is that a smaller basis set size could be used, reducing the computational cost, while maintaining acceptable accuracy.

ZSs can be viewed as Fermionic Coherent States (CS) of a two-level system.  Previously Coherent States of the harmonic oscillator have been used to describe Bosonic systems in second quantisation with the help of Herman-Kluck \cite{ISI:000372195600015} and Coupled Coherent States propagation methods \cite{PhysRevA.100.013607} and also with Generalised Coherent States\cite{PhysRevA.103.042209}. Second quantisation Hamiltonians look like those of coupled oscillators with the difference that the oscillators represent the amplitudes and populations of the orbitals occupied by Bosons.  Following the idea of these methods it seems reasonable to try to treat Fermions on a similar level.  The complication here is that the elements of Grassmann algebra enter into standard definition of a Fermionic coherent state\cite{alma1}: $\ket{\eta_j} = a(\ket{0_j}+\eta_j \ket{1_j})$ where $\ket{1_j}$ and $\ket{0_j}$ are the $j$-th orbital in its occupied and unoccupied vacuum states respectively, $a$ is a normalisation factor and $\eta_j$ is an element of Grassmann algebra. The necessity of Grassmann algebra makes computation difficult but is required to maintain the correct permutation antisymmetry of the multi-electron Fermionic Coherent States as well as the anticommutation of creation and annihilation operators. Zombie States were shown\cite{RN106} to be capable of describing fermionic coherent states while removing the need for Grassmann algebra and the use of Wick's theorem, which are normally required to evaluate matrix elements between Fermionic Coherent States\cite{wicks}.

Earlier work \cite{RN106} gave a mathematical treatment of the ZSs' second quantization for fermions, and creation and annihilation operators were defined with the use of a simple sign change rule, which replaces Grassmann algebra. These results were verified by the reproduction of Full Configuration Interaction\cite{RN111} electronic energies for Li$_2$ and LiH via diagonalization of the electronic structure Hamiltonian in the complete basis of randomly selected ZSs. It was also shown that Zombie States can be used for real time propagation of the electronic wave function. A Fourier transform of real-time Zombie State evolution also has reproduced the exact electronic energy levels in the above mentioned molecules.  Both methods are  however not very efficient and could be used only as a demonstration that Zombie States' mathematics was correct. It is therefore necessary to develop better ZSs-based methods for finding low-lying state electronic energies in molecules, which are most important in chemistry. In this vein it is necessary to improve upon the na\"ive algorithms initially presented to increase efficiency; computation of the Hamiltonian in the original article\cite{RN106} had $\mathcal{O}(M^5)$ scaling which can be reduced. This also follows for other key operators. Bosonic coherent states can be defined as an eigenstate of the annihilation operator whereas the only eigenstate of the annihilation operator for a fermionic coherent state is the vacuum. Thus we also seek to provide an alternative formalism for Zombie States in the language of second quantization that allows Zombie states to be created from a vacuum state. In this paper we will present the theoretical advancement in the formalism of Zombie states, deriving a stronger normalization condition. Efficient algorithms for key operators are then given utilising scaled algorithms and sensible manipulations of the program to greatly reduce the computation time of the Hamiltonian as well as other properties such as spin and the number of electrons. 

Further, imaginary time propagation is validated as an effective method for finding the ground-state energies of states of real Fermionic systems, capable of reproducing Full CI (in a truncated basis) \cite{RN111}, in this case for Li$_2$. We also show how a biasing method can be used to create a small basis of Zombie States, which with imaginary time evolution can find low-lying state energies. To improve further the accuracy we show how the procedure called "cleaning" removes the contribution of unwanted numbers of electrons and improves the accuracy of electronic energy. We also make an attempt to create a method for finding excited states. It has previously been shown that a Gram-Schmidt procedure can be applied to orthogonalize higher energy states against lower lying ones while adding minimal computational cost\cite{FCIQMC}. We apply a similar Gram-Schmidt procedure to imaginary time propagation of ZSs too showing low-lying excited states can be found in this manner.

This article is structured as follows: a brief recap of the original Zombie State formulation is given followed by the algebraic developments; the algorithmic improvements for the two-electron Hamiltonian are detailed; results for imaginary time evolution of Li$_2$ for various basis sets and finally results for imaginary time evolution for excited states of Li$_2$ using Gram-Schmidt orthogonalization.

\section{Theory}
\subsection{Original Formulation}\label{originalformulation}

Zombie states\cite{RN106} are coherent antisymmetrized superpositions of `dead' and `alive' electronic states. Considering the single $m$th spin orbital, a coherent state is
\begin{align}
 \ket{\zeta(a_{1m},a_{0m})} = a_{1m}\ket{1_m} + a_{0m} \ket{0_m}
\end{align}
where $\ket{1_m}$ corresponds to their being an electron in spin orbital $m$ and $\ket{0_m}$ to the $m$th spin orbital being empty, consistent with the conventional electronic structure notation.

We can generalise this to a coherent state which is a Slater Determinant of one-electron zombie states
\begin{align}
 \ket{\bm{\zeta}} = \ket{\zeta_1 \zeta_2 \ldots \zeta_M}
\end{align}
and which can be summarized by $2M$ coefficients
\begin{align}
 \ket{\bm{\zeta}} =  
 \begin{bmatrix}
 a_{11} & a_{12} & \ldots & a_{1(m-1)} & a_{1m} & a_{1(m+1)} & \ldots & a_{1M} \\
 a_{01} & a_{02} & \ldots & a_{0(m-1)} & a_{0m} & a_{0(m+1)} & \ldots & a_{0M} 
\end{bmatrix}
\end{align}
the notation used here is synonymous with that in Ref~\onlinecite{RN106}, where in $a_{m_j j}$, $m_j=0$ refers to the $m$th spin orbital being `dead' (unoccupied) and $m_j=1$ to the $m$th spin orbital being `alive' (occupied), and $j$ refers to the spin orbital number. Note that the zombie state contains $2M$ complex coefficients, where $M$ is the total number of spin orbitals.

The overlap of two states is
\begin{align}
 \Omega_{ab} = \bk{\bm{\zeta}^{(a)}}{\bm{\zeta}^{(b)}} = \prod_{j = 1}^{M}\sum_{m_j = 0,1} a_{m_j j}^{(a)*}a_{m_j j}^{(b)}. \eql{overlap}
\end{align}
Calculation of this is $\mathcal{O}(M)$ where there are $M$ spin orbitals, whether or not they are occupied.

The action of the creation and annihilation operators on a ZS is given by
\begin{subequations}
\begin{align}
\hat b^\dagger_m\ket{\bm{\zeta}^{(b)}} = &  
 \begin{bmatrix}
 -a_{11}^{(b)} & -a_{12}^{(b)} & \ldots & -a_{1(m-1)}^{(b)} & a_{0m}^{(b)} & a_{1(m+1)}^{(b)} & \ldots & a_{1M}^{(b)} \\
 a_{01}^{(b)} & a_{02}^{(b)} & \ldots & a_{0(m-1)}^{(b)} & 0 & a_{0(m+1)}^{(b)} & \ldots & a_{0M}^{(b)} 
\end{bmatrix}, \\
\hat b_m \ket{\bm{\zeta}^{(b)}} = &
 \begin{bmatrix}
 -a_{11}^{(b)} & -a_{12}^{(b)} & \ldots & -a_{1(m-1)}^{(b)} & 0 & a_{1(m+1)}^{(b)} & \ldots & a_{1M}^{(b)} \\
 a_{01}^{(b)} & a_{02}^{(b)} & \ldots & a_{0(m-1)}^{(b)} & a_{1m}^{(b)} & a_{0(m+1)}^{(b)} & \ldots & a_{0M}^{(b)} 
\end{bmatrix}. \eql{creatanihil}
\end{align}
\end{subequations}

The operators $\hat b^\dagger_m$ and $\hat b_m$ not only act on the $m$-th orbital but also change sign of the amplitudes $a_{1n}$ of alive coefficients for all orbitals with $n<m$. If $a_{1m}$ and $a_{0n}$ are given by ones and zeros so that Zombie States represent standard Slater Determinants, it is easy to see that the sign changing rule is equivalent to the so called Wigner-Jordan rule.  As previously shown all anticommutation relations are as expected \cite{RN106}. Most straightforwardly, but not very efficiently, matrix elements $\bra{\bm{\zeta}^{(a)}}\hat H\ket{\bm{\zeta}^{(b)}}$ of the second quantized electronic structure Hamiltonian can be computed by sequential application of the creation and annihilation operators\cite{RN106}

\begin{align}
\hat H=\sum_{m,n} h_{mn} \hat b^\dagger_m \hat b_n + \frac{1}{2} \sum_{klmn} \hat b^\dagger_k \hat b^\dagger_l W_{klnm} \hat b_m \hat b_n \eql{2ndquant}
\end{align}
to $\ket{\bm{\zeta}^{(b)}}$  and overlapping the result with $\bra{\bm{\zeta}^{(a)}}$,

\begin{align}
H_{ab}=\bra{\bm{\zeta}^{(a)}}\hat H \ket{\bm{\zeta}^{(b)}} = \sum_{m,n} h_{mn} \bk{\bm{\zeta}^{(a)}}{\bm{\zeta}_{mn}^{(b)}} +\dfrac{1}{2} \sum_{klmn} W_{klnm} \bk{\bm{\zeta}^{(a)}}{\bm{\zeta}_{klmn}^{(b)}},
\end{align}

where $\ket{\bm{\zeta}_{klmn}^{(b)}} =  \hat b^\dagger_k \hat b^\dagger_l \hat b_m \hat b_n \ket{\bm{\zeta}^{(b)}}$ and $\ket{\bm{\zeta}_{mn}^{(b)}} = b^\dagger_m \hat b_n \ket{\bm{\zeta}_{mn}^{(b)}}$; the overlaps are calculated using \eqr{overlap}.

An electronic wavefunction can be represented as a superposition of $K$ basis Zombie states 
\begin{align}
 \ket{\Psi} = \sum_k^K d_k \ket{\bm{\zeta}^{(k)}}.\eql{zsbasisone}
\end{align}
and the matrix elements described above now allow usage of the ansatz \eqr{zsbasisone} for propagation or finding quantum states and their energies.  Individual Zombie states generally are not restricted to a particular number of electrons, but for a wavefunction such as \eqr{zsbasisone} a sufficiently large $K$ number of zombie states with coefficients $d_k$ usually can be chosen such that contributions from unwanted numbers of electrons cancel out. This is very different from standard electronic structure methods, where the problem is to include as many configurations with the right number of electrons as possible.   

A standard Hartree Fock configuration $\ket{\varphi_{m_e}^{(j)}}$ which corresponds to $m_e$ electrons can be written as a Zombie State with "binary" amplitudes of dead and alive states and $m_e$ ones in the upper row.

\begin{align}
 \ket{\varphi_{m_e}^{(j)}}= 
 \begin{bmatrix}
 1 & 0 & \ldots & 0 & 1 & 0 & \ldots & 1 \\
 0 & 1 & \ldots & 1 & 0 & 1 & \ldots & 0 
\end{bmatrix} 
\eql{fihartreefock}
\end{align}
and it can be treated just like any other ZS. 

In the  subsections \ref{algb} and \ref{algorithm}  algebraic and algorithmic improvements are presented, which allow much more efficient calculation of matrix elements of operators between ZSs and faster propagation of the wave function \eqr{zsbasisone}. But a reader less interested in technical aspects of the method can go directly to the section \ref{imgnr}, where it is shown that the right electronic energies can be recovered from imaginary time propagation of the wave function \eqr{zsbasisone} with unrestricted number of electrons.  

\subsection{Algebraic Developments} \label{algb}
The previous article\cite{RN106} gave the action of creation and annihilation operators for zombie states and how the Hamiltonian could be computed for them. Here we extend the algebraic formalism by showing how a Zombie state can be created from a vacuum state. We show how creation and annihilation operators act in this formalism, as well as computation of the overlap of two Zombie states. Finally we derive a stronger condition for normalization of a Zombie state. 

From standard electronic structure theory\cite{RN111}, a given electronic occupancy can be written as an antisymmetrized Slater determinant
\begin{align}
 \ket{\varphi} = \ket{m_1 m_2 \ldots m_M}
\end{align}
where $\{m_i\} = 0$ if the $i$th orbital is empty and 1 if occupied, as detailed in Ref. \onlinecite{PhysRevA.81.022124}, or equivalently in the form \eqr{fihartreefock}. This is equivalent to stating in second quantization notation
\begin{align}
 \ket{\varphi} = \prod_{k \ \textrm{occ}} \hat b^\dag_k \ket{} \eql{prodcr}
\end{align}
where $\ket{}$ is the vacuum state, ensuring here and in what follows that the creation operators are applied in the reverse order they appear in the Slater determinant. We could also trivially rewrite \eqr{prodcr} as
\begin{align}
 \ket{\varphi} = \prod_{j=1}^{M} [(1-m_j) \hat I + m_j \hat b^{\dag}_j] \ket{}. \eql{prod}
\end{align}
We now consider a generalized form of \eqr{prod}
\begin{align}
 \ket{\bm{\zeta}} = \prod_{j=1}^{M} (a_{0j} \hat I + a_{1j} \hat b^{\dag}_j) \ket{}, \eql{proda}
\end{align}
where $a_{0j}$ and $a_{1j}$ are complex scalar coefficients. Clearly if $a_{1j} = m_j$ and $a_{0j} = (1-m_j)$ then the same Slater determinant electron state is produced as in \eqr{prod}.

By inference from \eqr{proda}, we can define a zombie operator
\begin{align}
 \hat z_j =: a_{0j} \hat I + a_{1j} \hat b^{\dag}_j,
\end{align}
whose adjoint is
\begin{align}
 \hat z_j^\dag =: a_{0j}^* \hat I + a_{1j}^* \hat b_j,
\end{align}
and commutators and anti-commutators 
\begin{subequations}
\begin{align}
&[\hat z_j, \hat z_k] = 2\hat b^\dag_j \hat b^\dag_k a_{1j} a_{1k}\\
& [\hat z_j^\dag, \hat z_k^\dag] = 2\hat b_j \hat b_k a_{1j} a_{1k}\\
 &\lbrace \hat z_j, \hat z_k\rbrace = 2(a_{0j}a_{0k}\hat I+a_{1j}a_{0k} \hat b_j^\dag +a_{0j}a_{1k} \hat b_k^\dag) \hat I\\
 &\lbrace \hat z_j^\dag, \hat z_k\rbrace = 2(a_{0j}a_{0k}\hat I+a_{1j}a_{0k} \hat b_j +a_{0j}a_{1k} \hat b_k),  
\end{align}
\end{subequations}
such that \eqr{proda} can be written more concisely as
\begin{align}
    \ket{\bm{\zeta}} = \prod_{j=1}^{M} \hat z_j  \ket{}. \eql{prodz}
\end{align}
Note that $\hat z_j$ is a function of two complex numbers, so we can write $\hat z_j \equiv \hat z_j(a_{0j},a_{1j})$. This shorthand will become useful later.

The idea of defining new operators from linear combinations of creation and annihilation operators is of course nothing new, the most famous example perhaps being Majorana Fermions, which are their own antiparticle, defined as
\begin{subequations} 
\begin{align}
 \gamma_1 = & \frac{1}{\sqrt{2}}(\hat b + \hat b^\dag), \\
 \gamma_2 = & \frac{1}{\sqrt{2}i}(\hat b - \hat b^\dag).
\end{align}
\end{subequations}
However, zombie operators are not their own antiparticles. We now consider the action of creation and annihilation operators on zombie operators.

\subsubsection{Creation and annihilation operators}
The action of creation and annihilation operators on a single zombie operator is given, for $j\neq k$, by
\begin{subequations}
\begin{align}
 \hat b_j \hat z_k(a_{0k},a_{1k}) = & \hat b_j (a_{0k} \hat I + a_{1k}\hat b^\dag_k) = (a_{0k} \hat I - a_{1k}\hat b^\dag_k)\hat b_j = \hat z_{k}(a_{0k},-a_{1k}) \hat b_j, \\
 \hat b^\dag_j \hat z_k(a_{0k},a_{1k}) = & \hat b^\dag_j (a_{0k} \hat I + a_{1k}\hat b^\dag_k) = (a_{0k} \hat I - a_{1k}\hat b^\dag_k)\hat b^\dag_j = \hat z_{k}(a_{0k},-a_{1k}) \hat b^\dag_j,
\end{align}
\end{subequations}
and if $j=k$,
\begin{align}
 \hat b_j \hat z_j(a_{0j},a_{1j}) = & \hat b_j (a_{0j} \hat I + a_{1j}\hat b^\dag_j) = a_{0j} b_j - a_{1k}(1-\hat b^\dag_j\hat b_j) = \hat z_j(a_{0j},-a_{1j}) \hat b_j + \hat z_j(a_{1j},0), \\
 \hat b^\dag_j \hat z_j(a_{0j},a_{1j}) = & \hat b^\dag_j (a_{0j} \hat I + a_{1j}\hat b^\dag_j) = a_{0j}\hat b^\dag_j = \hat z_{j}(0,a_{0j}),
\end{align}
where we use the standard relations\cite{RN111} $\{\hat b_j,\hat b_k\} = 0$, $\{\hat b^\dag_j,\hat b^\dag_k\} = 0$, and $ \{\hat b_j, \hat b^\dag_k\} = \delta_{jk}$.

Using these results we can compute the action of the creation and annihilation operators on a zombie state, which is defined by the product of zombie operators as in \eqr{prodz},
\begin{subequations}
\begin{align}
 \hat b_j \ket{\bm{\zeta}} = & \hat b_j \prod_{k=1}^M \hat z_k(a_{0k},a_{1k}) \\
 = & \left[ \prod_{k=1}^{j-1} \hat z_k(a_{0k},-a_{1k}) \right]
 \left[\hat z_j(a_{0j},-a_{1j}) \hat b_j + \hat z_j(a_{1j},0) \right] \left[ \prod_{k=j+1}^M \hat z_k(a_{0k},a_{1k}) \right] \ket{} \\
 = & \left[ \prod_{k=1}^{j-1} \hat z_k(a_{0k},-a_{1k}) \right]
 \hat z_j(a_{1j},0)  \left[ \prod_{k=j+1}^M \hat z_k(a_{0k},a_{1k}) \right] \ket{} \eql{anz}
\end{align}
\end{subequations}
where the $\hat z_j(a_{0j},-a_{1j}) \hat b_j$ term vanishes because $\hat b_j$ tries to destroy an electron which is not there.

It also follows
\begin{subequations}
\begin{align}
 \hat b^\dag_j \ket{\bm{\zeta}} = & \hat b^\dag_j \prod_{k=1}^M \hat z_k(a_{0k},a_{1k}) \\
 = & \left[ \prod_{k=1}^{j-1} \hat z_k(a_{0k},-a_{1k}) \right]
 \hat z_j(0,a_{0j})  \left[ \prod_{k=j+1}^M \hat z_k(a_{0k},a_{1k}) \right] \ket{}. \eql{crz}
\end{align}
\end{subequations}
However, \eqr{anz} and \eqr{crz} themselves describe zombie states with changed coefficients to those of $\ket{\zeta}$. The zombie states they refer to agree exactly with the sign-changing rules presented in the original zombie states paper, though through more convoluted algebra. 

These results can equivalently be stated in terms of the $\{a_{n_j j}\}$ coefficients. As $\ket{\zeta}$ is a function of the zombie coefficients $\{a_{n_j j}\} = \mathbf{a}$ we can write $\ket{\zeta} \equiv \ket{\zeta(\mathbf{a})}$. Then let $\ket{\zeta^{(j)}(\mathbf{a}^{(j)})} = \hat b_j \ket{\zeta(\mathbf{a})}$. We then see that 
\begin{subequations}
 \begin{align}
  a_{0k}^{(j)} = & \left\{
  \begin{array}{cc}
   a_{0k} & k \neq j \\
   a_{1k} & k = j
  \end{array},
  \right. \\
   a_{1k}^{(j)} = & \left\{
  \begin{array}{cc}
   -a_{1k} & k < j \\
   0 & k = j \\
   a_{1k} & k > j
  \end{array}.
  \right.
 \end{align}
\end{subequations}
If $\ket{\bm{\zeta}^{(j)}(\mathbf{a}^{(j)})} = \hat b^\dag_j \ket{\bm{\zeta}(\mathbf{a})}$
\begin{subequations}
 \begin{align}
  a_{0k}^{(j)} = & \left\{
  \begin{array}{cc}
   a_{0k} & k \neq j \\
   0 & k = j
  \end{array},
  \right. \\
   a_{1k}^{(j)} = & \left\{
  \begin{array}{cc}
   -a_{1k} & k < j \\
   a_{0k} & k = j \\
   a_{1k} & k > j
  \end{array},
  \right.
 \end{align}
\end{subequations}
again in exact agreement with the earlier results.

\subsubsection{Overlap of two Zombie states}
We can use this formalism to compute the overlap of two zombie states, $\bk{\bm{\zeta}^{(a)}}{\bm{\zeta}^{(b)}}$, where both the bra and ket are formed from the action of $M$ zombie operators as in \eqr{prodz}. We consider a computation recursively, noting that the vacuum state is normalized, namely $\bk{}{} = 1$, and defining
\begin{align}
 \ket{\bm{\zeta}^{(b)}_j} = \prod_{k=j}^{M} \hat z_k \ket{} \eql{pbjd}
\end{align}
such that
\begin{subequations}
\begin{align}
 \ket{\bm{\zeta}^{(b)}_1} \equiv & \ket{\bm{\zeta}^{(b)}}, \\
 \ket{\bm{\zeta}^{(b)}_{M+1}} \equiv & \ket{}.
\end{align}
\end{subequations}
We therefore have
\begin{subequations}
\begin{align}
 \bk{\bm{\zeta}^{(a)}_j}{\bm{\zeta}^{(b)}_j} = & \bra{\bm{\zeta}^{(a)}_{j+1}} \hat z_j^{(a)*} \hat z_j^{(b)} \ket{\bm{\zeta}^{(b)}_{j+1}} \\
 = & \bra{\bm{\zeta}^{(a)}_{j+1}} 
 a_{0j}^{(a)*} a_{0j}^{(b)} \hat I + 
 a_{0j}^{(a)*} a_{1j}^{(b)} \hat b_j^\dag +
 a_{1j}^{(a)*} a_{0j}^{(b)} \hat b_j +
 a_{1j}^{(a)*} a_{1j}^{(b)} \hat b_j \hat b_j^\dag
 \ket{\bm{\zeta}^{(b)}_{j+1}} \eql{longeq} \\
 = & (a_{0j}^{(a)*} a_{0j}^{(b)} + a_{1j}^{(a)*} a_{1j}^{(b)}) \bk{\bm{\zeta}^{(a)}_{j+1}}{\bm{\zeta}^{(b)}_{j+1}}
\end{align}\eql{pajpbj}
\end{subequations}
The second term in \eqr{longeq} is of the form $\bra{\zeta^{(a)}_{j+1}}\hat b_j^\dag\ket{\zeta^{(b)}_{j+1}}$, and since neither $\bra{\zeta^{(a)}_{j+1}}$ nor $\ket{\zeta^{(b)}_{j+1}}$ contain any contributions from electron $j$, $\hat b_j^\dag$ acting to the left will try to destroy an electron which is not there, giving zero. Similar arguments show why the third term in \eqr{longeq} vanishes, and why the fourth term survives.

Combining \eqr{pbjd}--\eqr{pajpbj} gives
\begin{align}
 \bk{\bm{\zeta}^{(a)}}{\bm{\zeta}^{(b)}} = \prod_{j=1}^{M} (a_{0j}^{(a)*} a_{0j}^{(b)} + a_{1j}^{(a)*} a_{1j}^{(b)}) \eql{ov}
\end{align}
which is identical to \eqr{overlap}, which was obtained by different algebraic means in the original zombie paper\cite{RN106}.

\subsubsection{Normalization}
Clearly, if we wish a zombie state (the product of $M$ zombie operators) to be normalized, then the RHS of \eqr{ov} must equal unity when $\ket{\bm{\zeta}^{(a)}} = \ket{\bm{\zeta}^{(b)}}$. However, we may want to consider a more general case, where we may not necessarily have wavefunctions formed from a fixed $M$ number of zombie operators, but a variable number of zombie operators. 

Algebraically, if we require
\begin{align}
 \prod_{j=1}^{M} |a_{0j}^{(a)}|^2 + |a_{1j}^{(a)}|^2 = 1, \eql{weaknorm}
\end{align}
this will ensure that $\bk{\bm{\zeta}^{(a)}}{\bm{\zeta}^{(a)}} \equiv 
\bk{\bm{\zeta}^{(a)}_1}{\bm{\zeta}^{(a)}_1} = 1$, but what if we further require
\begin{align}
 \bk{\bm{\zeta}^{(a)}_j}{\bm{\zeta}^{(a)}_j} = 1 \eql{strict}
\end{align}
for all $j$, $1 \leq j \leq M$?

This problem can also be solved recursively, since trivially $\bk{\bm{\zeta}^{(a)}_{M+1}}{\bm{\zeta}^{(a)}_{M+1}} \equiv \bk{}{}=1$. For $\bk{\bm{\zeta}^{(a)}_{M}}{\bm{\zeta}^{(a)}_{M}} = 1$, from \eqr{pajpbj}, it is required that $|a_{0M}^{(a)}|^2 + |a_{1M}^{(a)}|^2 = 1$. But from \eqr{pbjd}, $\bk{\bm{\zeta}^{(a)}_{j}}{\bm{\zeta}^{(a)}_{j}} = (|a_{0j}^{(a)}|^2 + |a_{1j}^{(a)}|^2) \bk{\bm{\zeta}^{(a)}_{j+1}}{\bm{\zeta}^{(a)}_{j+1}}$, and so for \eqr{strict} to be true for all $j$, we require
\begin{align}
 |a_{0j}^{(a)}|^2 + |a_{1j}^{(a)}|^2 = 1  \ \forall j \eql{strongestnorm}
\end{align}
Note that \eqr{strongestnorm} is a stronger criterion than \eqr{weaknorm}, i.e. satisfying \eqr{strongestnorm} will ensure that \eqr{weaknorm} is satisfied, but not the other way around.

\eqr{strongestnorm} might correspond to normalization of a "one dimensional" ZS. This normalization is equivalent to the normalization used when introducing quantum superposition sampling to the Multiconfigurational Ehrenfest method presented in Ref.~\onlinecite{RN122}. 

In the event that the zombie state is real (useful for imaginary time propagation to find stationary states) then a simple choice to satisfy \eqr{strongestnorm} is
\begin{align}
 a_{0j}^{(a)} = \cos(\theta_j), \quad a_{1j}^{(a)} = \sin(\theta_j) \eql{normcon}
\end{align}
where $\theta_j$ is a real number such that $0 \le \theta_j < 2\pi$. This means that a real, normalized zombie state can be completely described by $M$ real $\{\theta_j\}$ values. This can be generalized to complex states\cite{RN122} with 
\begin{align}
 a_{0j}^{(a)} = \cos(\theta_j), \quad a_{1j}^{(a)} = \sin(\theta_j)e^{i\phi_j \eql{randnormzom}}
\end{align}
where $\phi_j$ is drawn from the interval $0 \le \phi_j < 2\pi$.

\subsection{Algorithmic Developments} \label{algorithm}
Here we consider how to compute efficiently the action of the Hamiltonian and other operators on a zombie state. We show how the electronic Hamiltonian matrix elements with Zombie states $\bra{\bm{\zeta}^{(a)}} \hat H \ket{\bm{\zeta}^{(b)}}$ can be calculated more efficiently. Firstly we show how a more efficient algorithm with a smaller prefactor can calculate matrix elements faster and secondly we will present an algorithm with $\mathcal{O}(M^4)$ scaling as opposed to the $\mathcal{O}(M^5)$ scaling of the na\"ive algorithm used in the initial paper \cite{RN106} described above.

\subsubsection{Reduced prefactor Hamiltonian algorithms}
The bottleneck in the computation of FCI states and energies in general is evaluation of the two-electron Hamiltonian 
\begin{equation}
 \hat H_2 =  \frac{1}{2}\sum_{i,j,k,l} \bk{ij}{kl}\hat b^\dag_i \hat b^\dag_j \hat b_l \hat b_k \eql{h2}
\end{equation}
It was previously shown\cite{RN106} that the two-electron part of \eqr{2ndquant} can be evaluated as
\begin{align}
 \bra{\bm{\zeta}^{(a)}} \hat H_2 \ket{\bm{\zeta}^{(b)}} = \frac{1}{2}\sum_{ijkl}^M \bk{ij}{kl} \bk{\bm{\zeta}^{(a)}}{\bm{\zeta}^{(b)}_{ijlk}} 
 \eql{eq13}
\end{align}
where $\ket{\bm{\zeta}^{(b)}_{ijlk}} = b^\dag_i \hat b^\dag_j \hat b_l \hat b_k\ket{\bm{\zeta}^{(b)}}$. This can be solved with iterative applications of the creation and annihilation operators, meaning that the original algorithm  therefore requires $M^4$ terms to be evaluated to solve \eqr{eq13}. Each term has four creation/annihilation operators (each of which is $\sim \mathcal{O}(M)$ due to the sign changing rule) and one overlap computation, which is $\mathcal{O}(M)$ as
\begin{align}
 \Omega_{ab} = \bk{\bm{\zeta}^{(a)}}{\bm{\zeta}^{(b)}} = \prod_{m = 1}^{M}\sum_{n_m = 0,1} a_{n_m m}^{(a)*}a_{n_m m}^{(b)}. \eql{overlap2}
\end{align}
Consequently na\"ive evaluation of \eqr{2ndquant} requires $M^4$ creation/annihilation operations and is $\mathcal{O}(M^5)$ overall.

Many of the two-electron integrals are zero by spin symmetry, i.e. $\bk{ij}{kl} = 0$ unless $\sigma(i) = \sigma(k)$ and $\sigma(j) = \sigma(l)$, where $\sigma(i)$ is the spin of spin orbital $i$. One method for accelerating the na\"ive algorithm is by evalutating only $\ket{\bm{\zeta}^{(b)}_{ijlk}} = b^\dag_i \hat b^\dag_j \hat b_l \hat b_k\ket{\bm{\zeta}^{(b)}}$ if $\bk{ij}{kl} \neq 0$ on spin symmetry grounds. When applied to the restricted Hartree Fock calculation this simple code modification gave nearly a four times speed up when calculating a matrix element between two ten-orbital ZSs. This is roughly to be expected since three quarters of the two-electron integrals are zero from spin symmetry.

Further, it is also possible to loop only over combinations of $\{ijkl\}$ already known to be nonzero by altering the loops over orbitals such that only nonzero (by spin) $\ket{\bm{\zeta}^{(b)}_{ijlk}}$ are considered in the first place. This produces a comparable, albeit slightly larger time improvement, to ignoring zero two electron integrals.

In an attempt, to accelerate the code further, we note that computing the action of $M^4$ creation and annihilation operators is expensive, and look to evaluate fewer of them. Re-examining \eqr{h2} and \eqr{eq13} we see 
\begin{align}
 \bra{\bm{\zeta}^{(a)}} \hat H_2 \ket{\bm{\zeta}^{(b)}} = \frac{1}{2}\sum_{ijkl}^M \bk{ij}{kl} \bk{\bm{\zeta}^{(a)}_{ij}}{\bm{\zeta}^{(b)}_{lk}} 
 \eql{bboth}
\end{align}
where
\begin{align}
 \bra{\bm{\zeta}^{(a)}_{ij}} = & \bra{\bm{\zeta}^{(a)}} \hat  b^\dag_i \hat b^\dag_j  = \left(\hat b_j \hat b_i \ket{\bm{\zeta}^{(a)}}\right)^\dag \\ 
 \ket{\bm{\zeta}^{(b)}_{lk}} = &  \hat b_l \hat b_k\ket{\bm{\zeta}^{(b)}}
\end{align}
We can then compute $\{ \ket{\bm{\zeta}^{(a)}_{ij}} \} \forall i,j$ and $\{ \ket{\bm{\zeta}^{(b)}_{lk}} \} \forall l,k$, which requires $M^2$ creation and annihilation operator applications while overall the Hamiltonian evaluation is $\mathcal{O}(M^5)$. Having precomputed $\{ \ket{\bm{\zeta}^{(a)}_{ij}} \}$ and $\{ \ket{\bm{\zeta}^{(b)}_{lk}} \}$ we can then evaluate \eqr{bboth}.

From the definition of the overlap integral in \eqr{overlap}, we see that if \emph{both} the dead and alive zombie amplitudes are zero for a given state, i.e. for the $m$th spin orbital 
\begin{align}
\ket{\bm{\zeta}^{(c)}} = 
\begin{bmatrix}
 a_{11}^{(c)} & a_{12}^{(c)} & \ldots & a_{1(m-1)}^{(c)} & 0& a_{1(m+1)}^{(c)} & \ldots & a_{1M}^{(c)} \\
 a_{01}^{(b)} & a_{02}^{(c)} & \ldots & a_{0(m-1)}^{(c)} & 0& a_{1(m+1)}^{(c)} & \ldots & a_{0M}^{(c)}
\end{bmatrix}
\end{align}
then the overlap of this state with any other state will be zero. Furthermore, since the zombie creation and annihilation operators only move the dead and alive amplitudes within a spin orbital, and not between spin orbitals, $\ket{\bm{\zeta}^{(c)}}$ will continue to have zero overlap with any other state \emph{even after application of creation and annihilation operators}. Consequently if a state such as $\ket{\bm{\zeta}^{(c)}}$ is itself generated by application of creation and annihilation operators it will not contribute to the two-electron Hamiltonian and so does not need to be considered in further calculations.

We know that\cite{RN106}
\begin{align}
 \hat b \ket{\zeta}  \equiv \hat b
 \begin{bmatrix}
  a_1 \\ a_0
 \end{bmatrix}
 = \begin{bmatrix}
    0 \\ a_1
   \end{bmatrix}.
\eql{vanish}
\end{align}
Consequently, if $a_1 = 0$, application of $\hat b$ returns a vanishing state which will have no overlap with any other. This is a rephrasing, in zombie terms, of the well known result from electronic structure theory that trying to annihilate an electron which is not there will return zero. Applied to calculation of the two-electron Hamiltonian, let us consider $\ket{\bm{\zeta}^{(a)}_{ij}} =  \hat b_j \hat b_i \ket{\bm{\zeta}^{(a)}}$. If $a_{1i}^{(a)} = 0$, then $\hat b_i \ket{\bm{\zeta}^{(a)}}$ returns a vanishing state and will contribute nothing to the two-electron calculation, even after application of $\hat b_j$. If $\hat b_i \ket{\bm{\zeta}^{(a)}} =:  \ket{\bm{\zeta}^{(a)}_i}$ is nonzero, then if $a_{1j}^{(a)} = 0$, the zombie state vanishes. Consequently, when looping over orbital indices in \eqr{bboth}, if the application of a given annihilation operator would return a vanishing state (which can be determined from $a_{1i}^{(a)} = 0$) and similarly for $j,l,k$ then the all further calculation for that creation/annihilation operator can be discarded.

In addition, considering the overlap calculation \eqr{overlap} and rewriting it as 
\begin{align}
 \Omega_{ab} = \bk{\bm{\zeta}^{(a)}}{\bm{\zeta}^{(b)}} = \prod_{m = 1}^{M}  a_{0 m}^{(a)*}a_{0 m}^{(b)} + a_{1 m}^{(a)*}a_{1 m}^{(b)} \eql{overlaptwo}
\end{align}
we see that if \emph{any} of the $a_{0 m}^{(a)*}a_{0 m}^{(b)} + a_{1 m}^{(a)*}a_{1 m}^{(b)}$ terms are zero, then $\Omega_{ab} = 0$. This means that calculation can be terminated as soon as one of the terms is zero and the overlap returned as zero, without having to compute any remaining terms. These improvements gave a similar time improvement to previous changes when working with ten orbitals, however, the time improvement was approximately half as good when calculating a matrix element with 50 orbitals. Full details of the algorithmic code racing can be found in \textit{Section 1.1} of the supplementary material. All of these algorithms still have (approximate) fifth order scaling, i.e. $\mathcal{O}(M^5)$, although with a much lower prefactor than earlier versions.

\subsubsection{Lower-scaling Hamiltonian algorithm}
It is, however, possible to reduce the scaling of the two-electron Hamiltonian from $\mathcal{O}(M^5)$ to $\mathcal{O}(M^4)$. We firstly note that evaluation of the two-electron Hamiltonian could be written as
\begin{align}
 \bra{\bm{\zeta}^{(a)}} \hat H_2 \ket{\bm{\zeta}^{(b)}} = \frac{1}{2}\sum_{ijkl}^M \bk{ij}{kl} \bra{\bm{\zeta}^{(a)}_{ij}} \hat b_l \ket{\bm{\zeta}^{(b)}_{k}}. 
 \eql{ij_l}
\end{align}
We have previously shown that $\bra{\bm{\zeta}^{(a)}_{ij}}$ can be calculated in $\mcob{M^3}$ and $\ket{\bm{\zeta}^{(b)}_{k}}$ in $\mcob{M^2}$, and these can be calculated separately. Using previous methods, even with the calculation improvements, to evaluate $\bra{\bm{\zeta}^{(a)}_{ij}} \hat b_l \ket{\bm{\zeta}^{(b)}_{k}} $ $\forall i,j,k,l$ require $\mcob{M^3}$ for looping over $i,j,k$, then an $\mcob{M}$ for summing over $l$, then another $\mcob{M}$ for evaluating the overlap, making $\mathcal{O}(M^5)$ overall. We therefore consider how to evaluate $\bra{\bm{\zeta}^{(a)}_{ij}} \hat b_l \ket{\bm{\zeta}^{(b)}_{k}} $ for a specific set of $i,j,k$, but over all values of $l$, in $\mcob{M}$ operations, such that the overall algorithm would be $\mcob{M^4}$.

We start by simplifying our notation to $\bra{\bm{\zeta}^{(a)}} \hat b_l \ket{\bm{\zeta}^{(b)}} $ which yields
\begin{align}
 \bra{\bm{\zeta}^{(a)}} \hat b_l \ket{\bm{\zeta}^{(b)}} = 
 \left( \prod_{i=1}^{l-1} 
 a_{0 i}^{(a)*}a_{0 i}^{(b)} - a_{1 i}^{(a)*}a_{1 i}^{(b)}\right)
 \cdot 
 a_{0 l}^{(a)*}a_{1 l}^{(b)} 
 \cdot
 \left( \prod_{j=l+1}^{M} 
 a_{0 j}^{(a)*}a_{0 j}^{(b)} + a_{1 j}^{(a)*}a_{1 j}^{(b)}\right)
\end{align}
Note the minus sign in the first product caused by the action of the annihilation operator. We then define
\begin{subequations}
\begin{align}
 s_i = & a_{0 i}^{(a)*}a_{1 i}^{(b)},\eql{ham_low1} \\
 e_i = & a_{0 i}^{(a)*}a_{0 i}^{(b)} - a_{1 i}^{(a)*}a_{1 i}^{(b)},\eql{ham_low2} \\
 f_i = & a_{0 i}^{(a)*}a_{0 i}^{(b)} + a_{1 i}^{(a)*}a_{1 i}^{(b)}.\eql{ham_low3}
\end{align}

\end{subequations}
These can all trivially be calculated in $\mcob{M}$ steps.
We also define
\begin{subequations}
\begin{align}
 g_l = & \prod_{i=1}^{l} e_i, \eql{ham_low6}\\
 h_l = & \prod_{i=l}^{M} f_i \eql{ham_low7}
\end{align}
\end{subequations}
which can be calculated recursively in $\mcob{M}$ steps. Then
\begin{align}
 \bra{\bm{\zeta}^{(a)}} \hat b_l \ket{\bm{\zeta}^{(b)}} = g_{l-1} s_l h_{l+1}
\end{align}
which can also be found in $\mcob{M}$ steps.

The reduced scaling improves calculation of two-electron matrix elements over 20 times for 10 orbitals calculation and over 100 times with 50 orbital calculations compared to the na\"ive calculation. Further details are given in \textit{Table 4} in the supplementary material.

We have also developed an algorithm to calculate the number operator, which scales linearly with the number of orbitals in the system. Using similar method we also show how to reduce the scaling of the algorithm calculate $S_z$, $\hat S_z^2$ and total spin.  Details of these two algorithms are presented in appendices A and B. 

\subsection{Imaginary time evolution, wave function cleaning, biasing, and excited state calculation} \label{imgnr}
Imaginary time propagation, i.e.\ solution of the Schr\"odinger equation in imaginary time 
\begin{align}
 -\dd{}{\beta} \ket{\Psi} = \hat H \ket{\Psi} \eql{imtp}
\end{align}
is commonly used to find the ground state of a quantum mechanical system\cite{doi:10.1063/1.3193710, doi:10.1063/1.3302277, FCIQMC}. Many methods of Quantum Monte-Carlo exploit it in electronic structure theory. For example a random walk in the Fock space of Slater determinants has been shown to be a beautiful technique to solve \eqr{imtp} and to avoid the wave function node problem, which otherwise makes quantum Monte-Carlo for Fermions difficult.  In this section we will show that a small randomly-selected set of Zombie States can provide an accurate description of the ground state of a molecule.   

\subsubsection{Imaginary time algorithm}
Consider a wavefunction $\ket{\Psi}$ that can be expanded in the basis of  $K$ zombie states $\ket{\bm{\zeta}^{(k)}}$, where $K \le 2^{M}$, and which are normalised but not necessarily orthogonal,
\begin{align}
 \ket{\Psi} = \sum_k^K d_k \ket{\bm{\zeta}^{(k)}}.\eql{zsbasis}
\end{align}
Then applying the identity matrix from  Ref. \onlinecite{hubhel} for a nonorthogonal basis set $\sum_{l, k} \ket{\bm{\zeta}^{l)}} \Omega_{lk}^{-1}\bra{\bm{\zeta}^{(k)}}$ and using the reasoning from Ref. \onlinecite{SHAL04} we yield
\begin{align}
 \sum_k \frac{\partial{d_k}}{\partial \beta} \bk{\bm{\zeta}^{(l)}}{\bm{\zeta}^{(k)}} = -\sum_k \bra{\bm{\zeta}^{(l)}} \hat H \ket{\bm{\zeta}^{(k)}}d_k
 \eql{khd1}
\end{align}
which can be rearranged in matrix notation as
\begin{align}
 \mathbf{\dot d} = -\boldsymbol{\Omega}^{-1} \mathbf{Hd}. \eql{khd}
\end{align}
Note how the inverse overlap matrix with the elements $\boldsymbol{\Omega}_{lk}=\bk{\bm{\zeta}^{(l)}}{\bm{\zeta}^{(k)}}$ is required and how the algebra developed for Bosonic Coherent States is being applied to Fermionic ZSs.\cite{SHAL04}

\subsubsection{Cleaning}
The propagation of \eqr{khd} relaxes the wave function to the lowest energy state of the system\footnote{Providing that the starting wavefunction contains contributions from wavefunctions of the same symmetry and number of electrons as the lowest-energy state}. As occupations of ZSs spin-orbitals are fractional Zombie States and their linear combinations are not restricted to a particular number of electrons. However, they can be projected onto a Fock Space with a given number of electrons. The identity operator can be written as a sum of $m_e$ electron identities $\hat I_{m_e}$:
\begin{align}
\hat I= \sum_{m_e=0,M} \hat I_{m_e}, \eql{im}
\end{align}
where the identity covering the Fock Space with $m_e$ electrons is
\begin{align}
\hat I_{m_e}= \sum_{j} \ket{\varphi_{m_e}^{(j)}} \bra{\varphi_{m_e}^{(j)}}  \eql{identityme1}
\end{align}
where the sum in $j$ is over all Fock configurations $\varphi_{m_e}^{(j)}$ with the $m_e$ number of electrons, which can be viewed as Zombie states with "binary" amplitudes 1 and 0 and $m_e$ unit amplitudes of alive electrons on $m_e$ occupied spin-orbitals. 

Then the superposition of Zombie States in \eqr{zsbasisone} or \eqr{zsbasis} can be projected onto Fock Space of $m_e$ electrons as
\begin{align}
 \ket{\Psi_{m_e}} = \hat I_{m_e} \ket{\Psi} =  \sum_j c_j \ket{\varphi_{m_e}^{(j)}} \eql{zsbasis1}
\end{align}
where
\begin{align}
c_j= \sum_{k=1,N_{zs}}d_k \bk{\varphi^{(j)}}{\bm{\zeta}^{(k)}}.
\end{align}
The energy of the $m_e$ contribution and its norm can then be calculated as
\begin{align}
E_{m_e}= \sum_{ji} c_i^* c_j \bra{\varphi_{m_e}^{(i)}}\hat H \ket{\varphi_{m_e}^{(j)}},
\end{align}
and
\begin{align}
N_{m_e}= \sum_{j} c_j^* c_j ,
\end{align}
respectively.  The sum of energies of all $m_e$ contributions is equal to the energy of the whole Zombie wave function \eqr{zsbasisone} and all  $m_e$ norms add up to $1$, i.e.
\begin{align}
\bra{\Psi}\hat H \ket{\Psi}= \sum_{m_e}E_{m_e}
\end{align}
and
\begin{align}
\sum_{m_e}N_{m_e}=1.
\end{align}
For a completely converged Zombie wave function the energy and norm of all $m_e$ other than the right number of electrons $n_e$ will be zero, but if convergence is incomplete the energy of the converged $m_e$ wave function $\ket{\Psi_c}$ can be estimated as    
\begin{align}
\bra{\Psi_c}\hat H \ket{\Psi_c} \approx E_{m_e}/N_{m_e}.
\end{align}
by division of the $m_e$ energy by the $m_e$ norm. 

\subsubsection{Biased Bases}
In this section we will consider vaious ways of how a basis set of ZSs can be chosen to solve \eqr{khd}. The best choice (in theory) would be a complete set of $2^M$ Slater determinants $\ket{\varphi_{m_e}^{(j)}}$ covering all possible electron occupancies from 0 to $M$ electrons in $M$ spin-orbitals.  Diagonalization of the Hamiltonian matrix in this basis gives all possible energy levels which possible numbers of electrons from $0$ to $M$ which $M$ available spin orbitals can yield. A basis of randomly selected $2^M$ ZSs $\ket{\bm{\zeta}^{(k)}}$ is also complete and is capable of yeilding correct quantum energies and wave functions. However, for efficient calculation the basis set must be as small as possible.   We employ a biased basis which sits between the two extremes of complete order and randomness. In CASSCF calculations\cite{casscf} electrons and orbitals are split into three groups: inactive, active and virtual. From this we have designed a biasing method to set up a ZS basis. Inactive, or core electrons, are low lying and are always (at least in a CASSCF calculation) to be occupied; the active orbitals can either be occupied or unoccupied and virtual orbitals are always empty. We take this idea of splitting electrons into different groups and then exploit the fractional occupation of a Zombie state. This allows orbitals to be set as completely "alive" (or "dead") while also biasing orbitals to be more or less "dead" or "alive". Amplitudes for zombie states can be randomly chosen from normal distributions centred around either completely "alive" or completely "dead" values and the distributions' width set according to its activity level. To satisfy the normalization condition given by \eqr{strongestnorm}, in this paper, only one Zombie amplitude (dead or alive) is generated using the normal distribution and the other is set using \eqr{normcon}. So, the low-lying core electrons have "alive" amplitudes set to one, although a very narrow gaussian could also be employed, which corresponds to a normal distribution width of zero giving a $\delta$ function with $a_{1j}=1$ and $a_{0j}=0$. The active electron normal distributions can be centred to favour either completely alive or dead amplitudes; the point where centring changes from "alive" to "dead" naturally changes with each system. Thinner distributions are used for orbitals at either end of the active electron/orbital group and wider distributions for orbitals where the chance, or not, of occupation is relatively equal. Virtual orbitals would have "dead" coefficients set to 1. But again, narrow distribution near completely "dead" amplitudes can be employed.

\subsubsection{Excited States}
As previously stated the aim of developing Zombie states is so it can be implemented in no-adiabatic molecular dynamics simulations and so it would be useful to compute low-lying excited states with the low computational expense that Zombie states could provide. It has previously been shown that a Gram-Schmidt orthogonalization procedure can be combined with full configuration interaction quantum Monte Carlo\cite{FCIQMC} and was found to add little extra computational cost while allowing multiple low-energy states to be studied. We will apply the same process here combining imaginary time propagation and Gram-Schmidt. Generally Gram-Schmidt orthogonalization, for any set of vectors, is defined by
\begin{align}
 \mathbf{u}_k=\mathbf{v}_k-\sum_{j=1}^{k-1} \mathrm{proj}_{\mathbf{u}_j}(\mathbf{v}_k)
\end{align}
with the projection operator
\begin{align}
 \mathrm{proj}_u(\mathbf{v}) =\dfrac{\langle \mathbf{u}, \mathbf{v}\rangle}{  \langle \mathbf{u}, \mathbf{u}\rangle}.
\end{align}
In this implementation the $k$ lowest states of interest are arbitrarily set creating $\mathbf{d}_k$ vectors which are not necessarily orthogonal. The vectors are then each orthogonalized
\begin{align}
    \mathbf{d}_k^{'}=\mathbf{d}_k-\sum_{j=1}^{k-1} \mathrm{proj}_{\mathbf{d}^{'}_j}(\mathbf{d}_k),
\end{align}
where $\mathbf{d}^{'}_k$ denotes an orthogonalised vector, they are then normalised. Each state $\mathbf{d}^{'}_k$ is then put into the differential equation equivalent to \eqr{khd}
 \begin{align}
 \mathbf{\dot d}_k = -\Omega^{-1} \mathbf{H}\mathbf{d}_k^{'}.
\end{align}
A single time step is taken and a new set of $\mathbf{d}_k(\beta + \Delta \beta) =\mathbf{d}^{'}_k(\beta) +\mathbf{\dot d}_k(\beta) \Delta \beta$ are then found. The process then repeats: vectors are orthogonalised using the Gram-Schmidt process, normalised and a time step is taken. 

\section{Results}

Here we will demonstrate that imaginary time propagation is an effective and efficient method for finding ground state energies using a Slater determinant basis as a point of reference. Earlier work\cite{RN106} used Li$_2$ to verify the theoretical basis of Zombie states and so here we too will use Li$_2$ as an example system.  PyScf \cite{PYSCF} using a 6-31G$^{**}$ basis has been used to calculate one and two electron integrals. Five spatial molecular orbitals (and therefore 10 spin orbitals) and their one and two electron integrals were calculated. All calculations are carried out in atomic units so energies are in Hartrees. 

\subsection{Ground-state imaginary time propagation}

The first trivial case considered is a complete basis of Slater determinant $\ket{\varphi_{m_e}^{(j)}}$ Zombie states. Li$_2$ consists of 10 spin-orbitals which gives a full basis a size of $2^{10}=1024$ basis functions. In \figr{imgsld6} the six electron restricted Hartree Fock determinant is used as a starting point and imaginary time evolution gives the neutral ground state energy of Li$_2$. A full sized basis of $2^{10}=1024$  randomly generated basis functions $\ket{\bm{\zeta}^{(k)}}$ was also tested. Random zombie states are generated by randomly generating $\theta$, between $0$ and $2\pi$ for each orbital the dead and alive coefficients are then found by calculating $a_1=\cos(\theta)$ and $a_0=\sin(\theta)$ respectively. The random basis is made up of Zombie states consisting of superposition of "dead" and "alive" electrons compared to the Slater determinant basis which is entirely binary. The initial vector is a superposition of random zombie states which is chosen to be equal to the RHF determinant, and that propagation of this using imaginary time evolution (solid line in \figr{imgsld6}) matches the result obtained by using the Slater Determinant basis. The final energy of each state obtained through imaginary time propagation is verified by comparison to the eigenvalues found by diagonalizing the complete Slater determinant Hamiltonian.
\begin{figure}[H]
\centering
\includegraphics[width=\columnwidth]{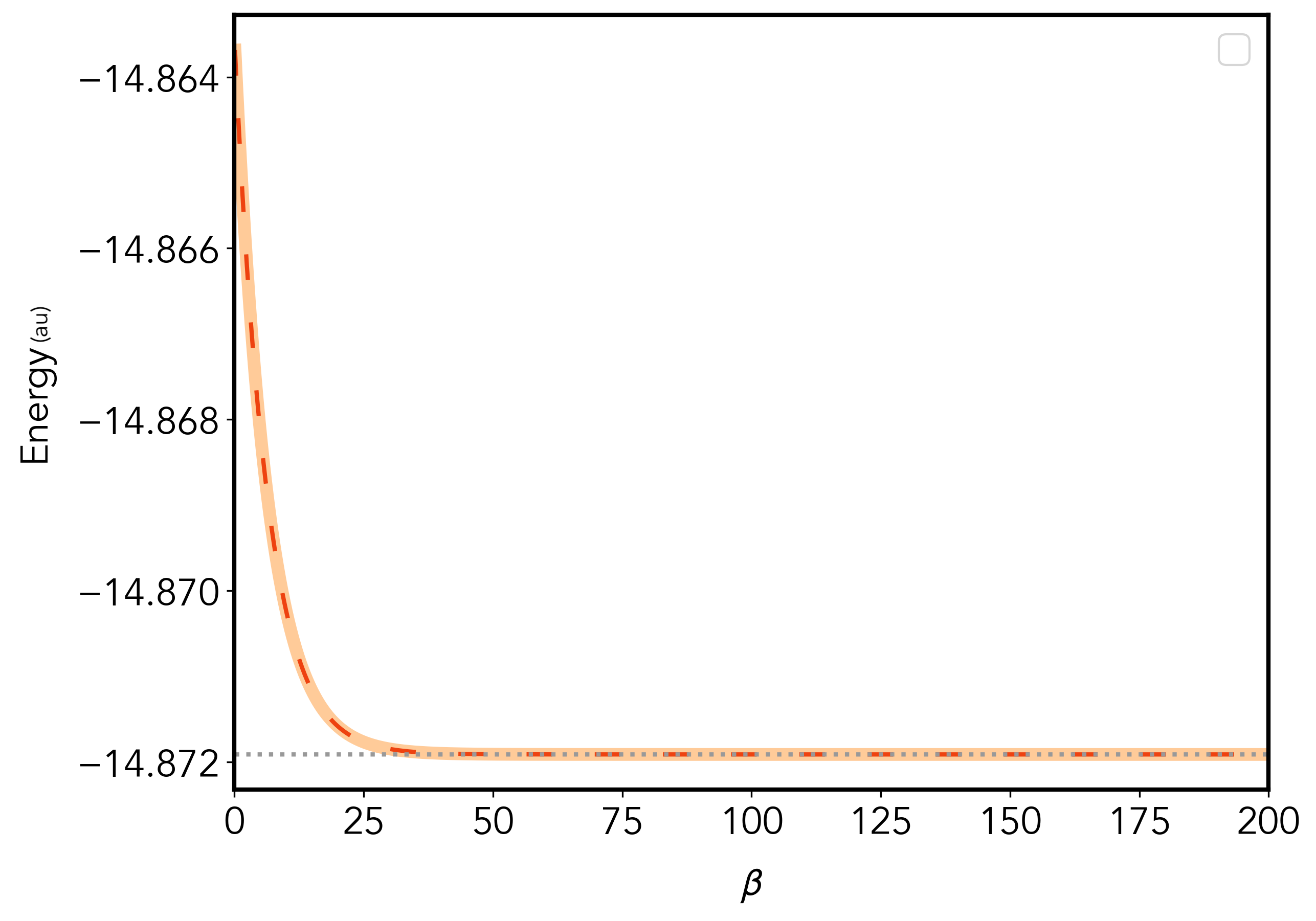}
 \caption{Imaginary time propagation starting with the 6-electron restricted Hartree Fock Slater determinant using a complete Slater determinant basis (dashed, red line) and a complete random basis (solid, orange line). Both results tend to neutral ground sate energy of Li$_2$ found by diagonalizing the Slater determinant Hamiltonian, shown as a  grey dotted line.}
\figl{imgsld6}
\end{figure}

\begin{figure}[H]
    \centering
    \begin{subfigure}{\columnwidth}
    \centering
    \includegraphics{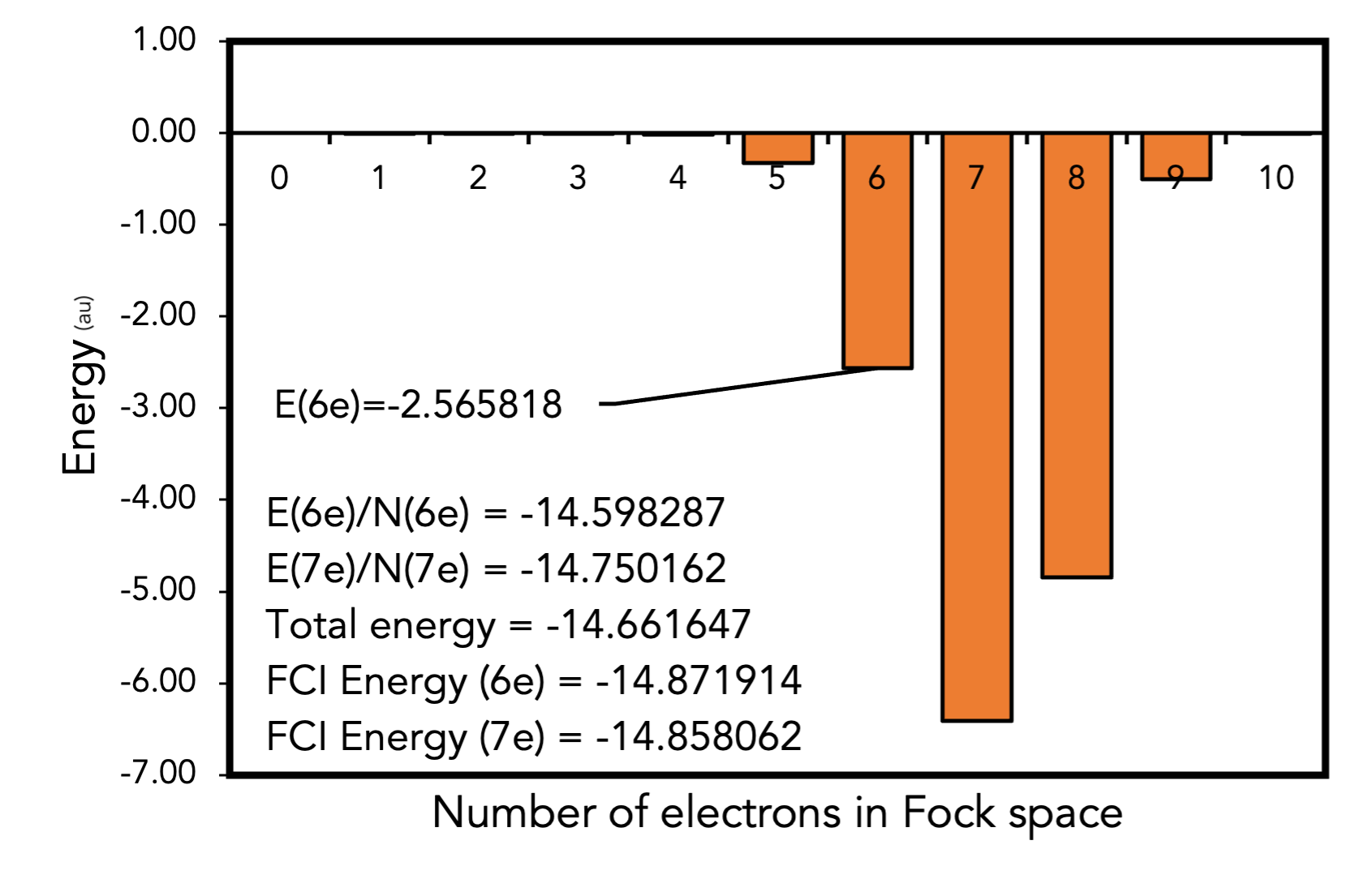}
    \end{subfigure}
    \begin{subfigure}{\columnwidth}
    \centering
    \includegraphics{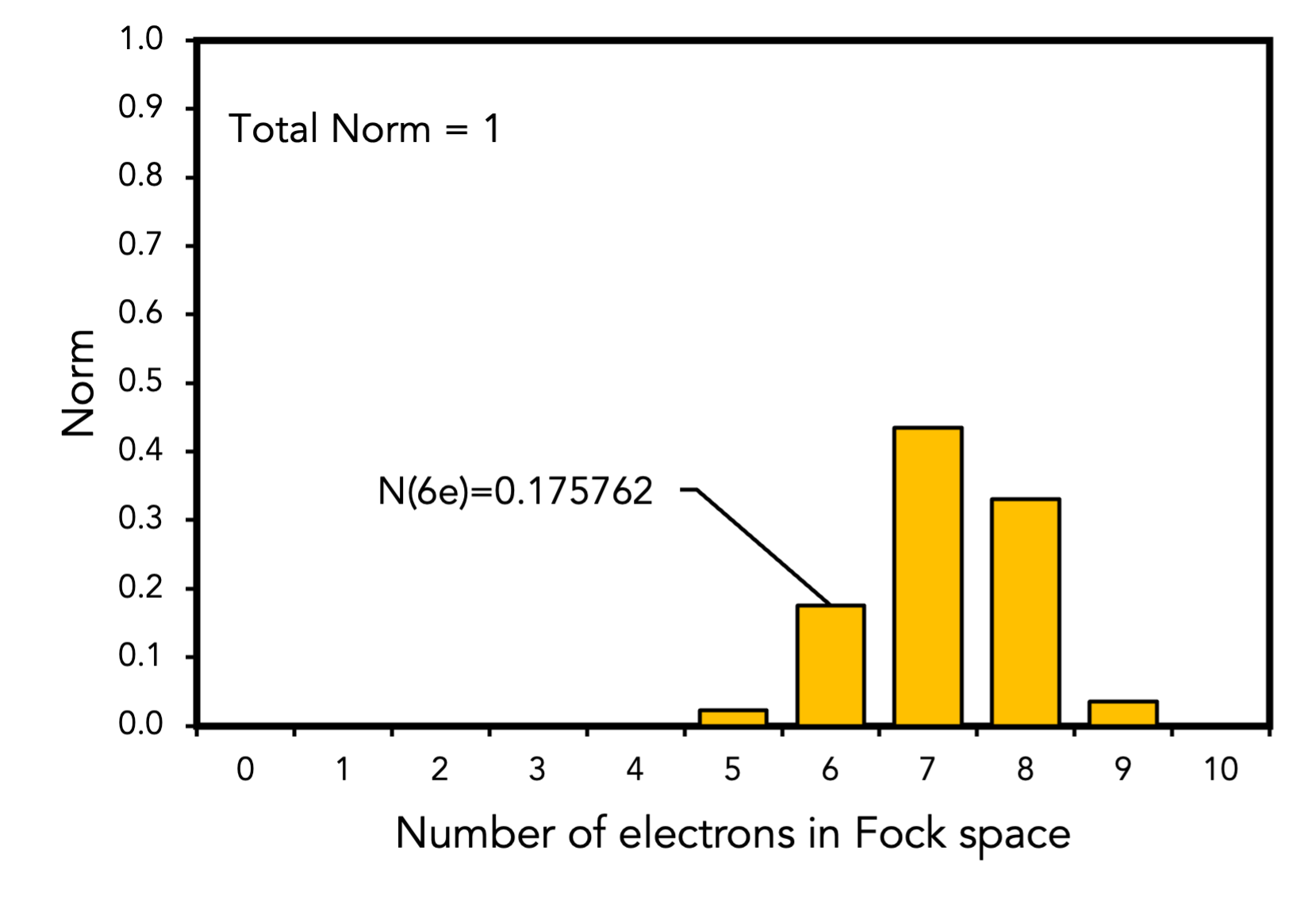}
    \end{subfigure}
    \caption{Energy (top) and norm (bottom) distribution for each number of electrons for a basis set of 200 randomly generated Zombie states. Only a fraction of the final wave function belongs to the Fock space on $n_e=6$ electrons.}
    \label{fig:rand200clean}
\end{figure}

When a reduced basis of 200 random Zombie states is used as seen in \figr{img200rand}, the basis set is no longer capable of reproducing the ground state energy. The reduced basis is created by selecting 200 basis functions $\ket{\bm{\zeta}^{(k)}}$ at random from the complete random basis previously used. The energy obtained via imaginary time propagation, in this smaller basis, no longer matches the value found by diagonalising the Slater determinant basis. But the cleaned energy provides a much better estimate of the ground state energy.  For a poorly selected basis the distribution of energy and norm over possible electronic numbers $m_e$ is shown in \figr{rand200clean}, which demonstrates that only a fraction  $(\approx17.5\%)$ of the final wave function belongs to the right Fock space on $n_e=6$ electrons. The highest fraction of the norm $(\approx40\%)$ belongs to the 7e Fock space.  The cleaned energy $E(6e)/N(6e)=-14.598267$ is somewhat above the FCI energy of the $6e$ ground state $(-14.871914)$, but the cleaned energy of the $7e$ Fock Space $E(7e)/N(7e)=-14.750162$ is approaching the $7e$ FCI $(-14.858062)$ energy. Thus the randomly selected basis has, by chance, been set up in favor of the anion.   

\begin{figure}[H]
\centering
\includegraphics[width=\columnwidth]{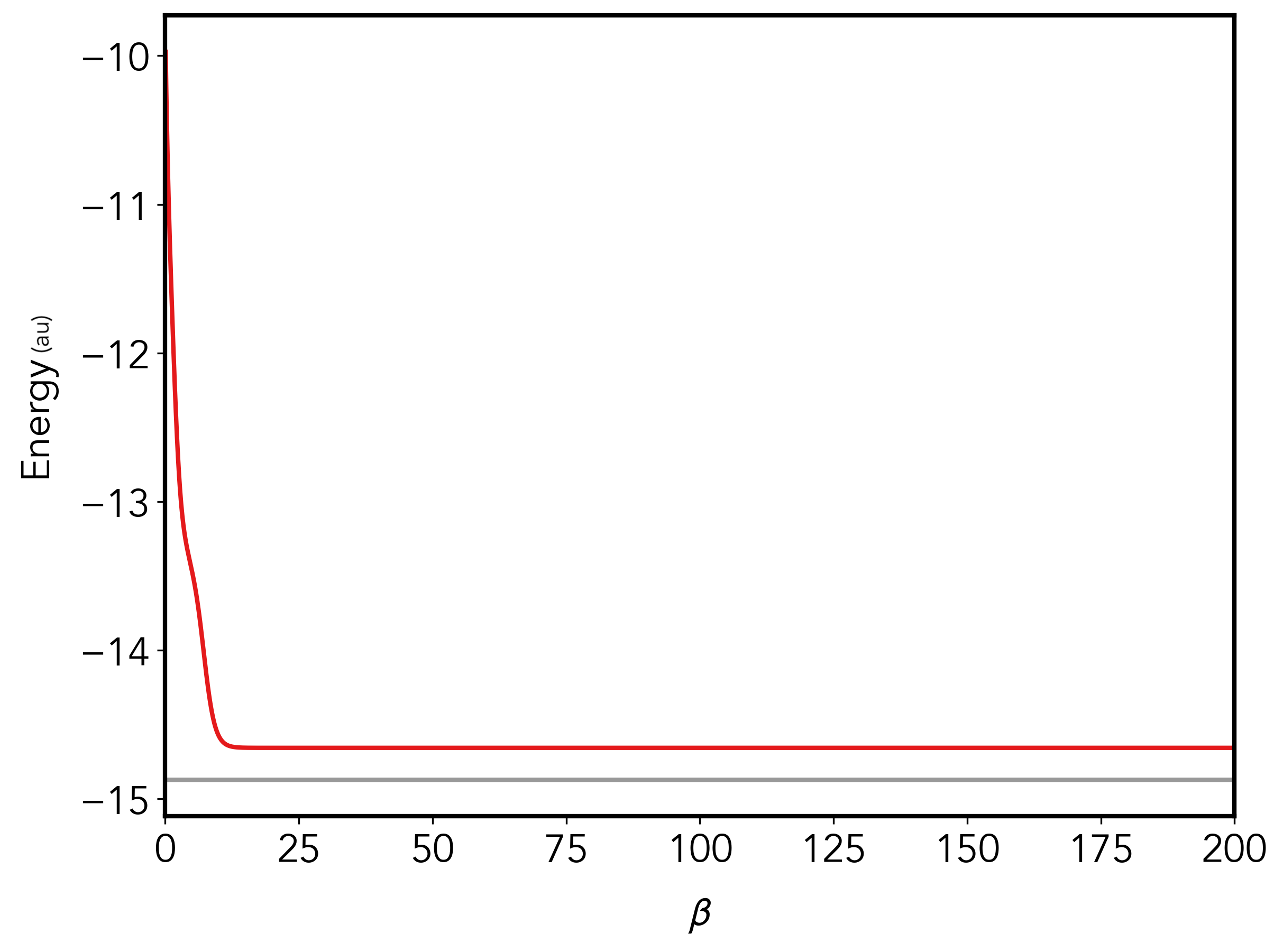}
\caption{Imaginary time propagation starting with the 6 electron restricted Hartree Fock Slater determinant and using a reduced random basis of 200 Zombie sates. The neutral ground sate energy of Li$_2$ found by diagonalising the Slater determinant Hamiltonian is also shown in grey. }
\figl{img200rand}
\end{figure}

The electronic structure of the Li$_2$ is not completely random and unknown so it is sensible to exploit this. A biased random basis set of Zombie states was generated using the biasing regime previously described. \eqr{randnormzom} was used to ensure the "dead" and "alive" coefficients were normalised, values $\theta_j$ were randomly generated using a normal distribution centred at $\dfrac{1}{2}\pi$ for electrons "alive" in the restricted Hartree Fock Slater determinant [$j=1\dots 6$] and centred around 0 for the "dead" electrons. The width, $\sigma$, of the normal distributions used to bias each electron are summarised in Table \ref{table:biased} and these values are used to create the biased bases in all subsequent results.There are six electrons in the neutral Li$_2$ molecule and four core inactive electrons which are set as always "alive".  \figr{biasedocc} schematically shows how this biasing method is applied: the first two spatial orbitals (first four spin orbitals) are set to always be occupied with the final three spatial orbitals (last six spin orbitals) having fractional occupancy. \begin{table}[H]
\centering
\begin{tabular}{lll}
j-th Electron & \multicolumn{2}{l}{Figure} \\
              & $\mu$  $\setminus2\pi$        &$\sigma$  $\setminus2\pi$         \\
\hline          
1, 2, 3, 4    & 0.25           &0           \\
5, 6          & 0.25           &0.175           \\
7, 8          & 0	          &0.351           \\
9, 10         & 0              &0.120          
\end{tabular} 
\caption{The centre $\mu$ and width $\sigma$ of the normal distributions used to generate $\theta_j$ for the $j$-th electron of each Zombie state when creating a biased basis.}
\label{table:biased}
\end{table}

\begin{figure}[H]
\centering
\includegraphics[width=\columnwidth]{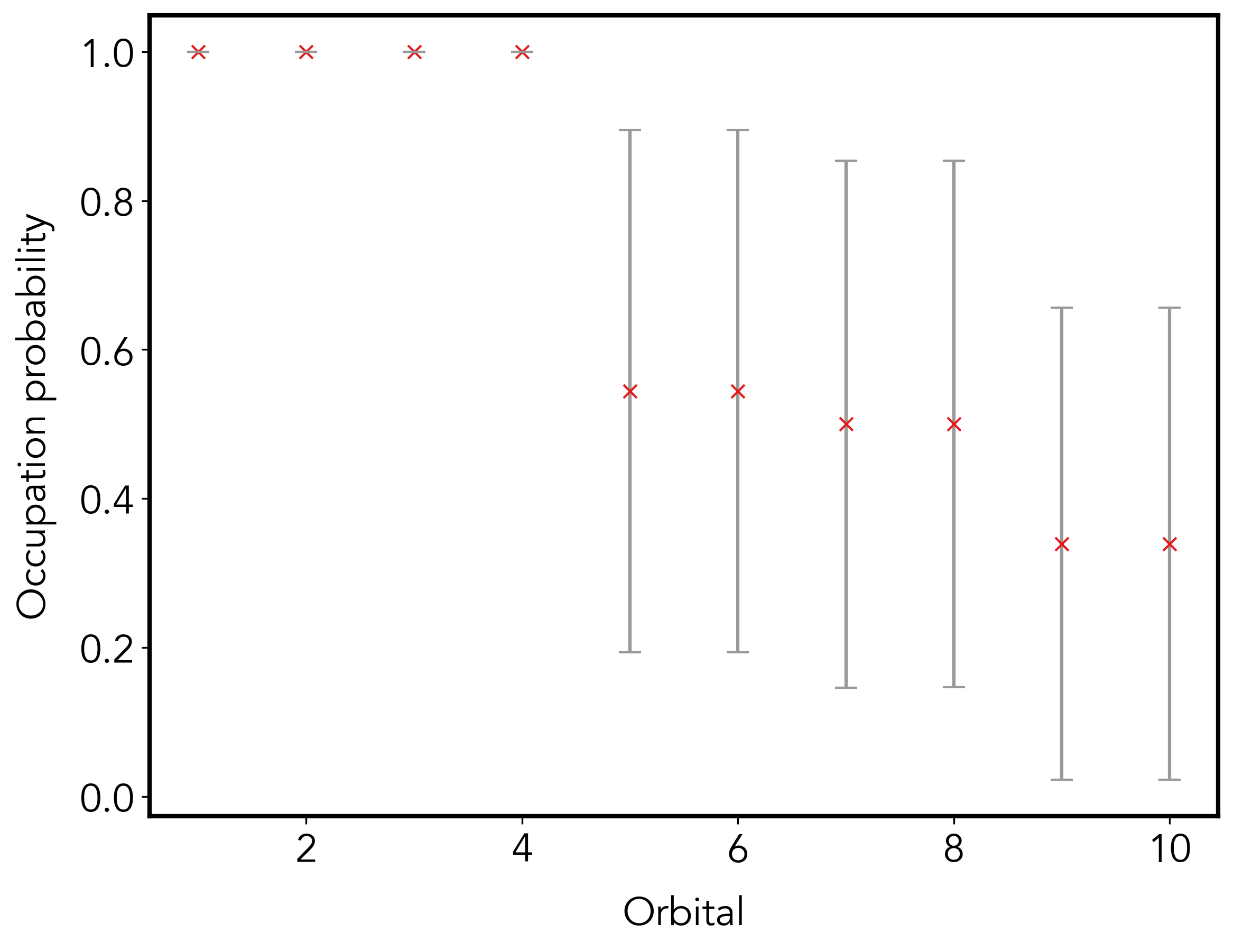}
\caption{Occupational probability plotted against orbital number for five spin orbitals. The first four spin orbitals are set exactly as occupied, the rest are set using the Gaussian described in table \ref{table:biased}. The average probability is shown by a cross and one standard deviation either side is the bar.}
\figl{biasedocc}
\end{figure}

\begin{figure}[H]
\centering
 \includegraphics[width=\columnwidth]{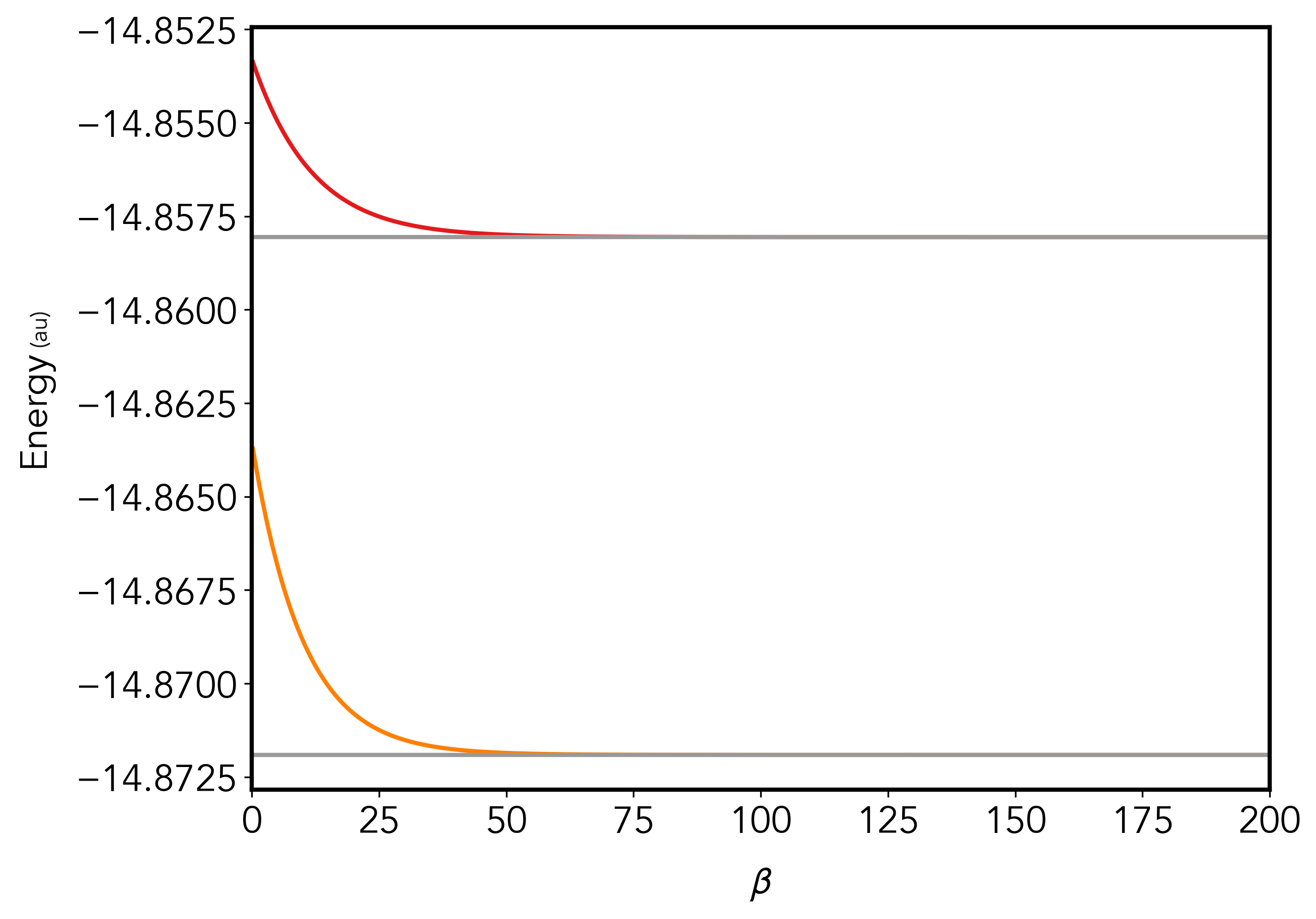}
 \caption{Imaginary time propagation for two biased bases each made up of 63 biased Zombie state basis functions and then a Zombie state set either to the 6 electron restricted Hartree Fock determinant (orange bottom line) or the 7 electron open-shell restricted Hartree Fock determinant (red top line). The basis with six electron determinant evolves to the Li$_2$ neutral ground state and the basis with seven electrons evolves to the higher $Li_2^-$ anion ground state. The corresponding energies, obtained by diagonalising the complete Slater determinant basis, are also shown as horizontal lines. The details of how each electron is biased are summarised in Table \ref{table:biased}.}
\figl{imgbiased64HF67}
\end{figure}

\figr{imgbiased64HF67} also demonstrates that ZS propagation can yield the ground state of the  Li$_2^-$ anion. A 64 biased Zombie state basis was generated but the first basis function was set as the seven electron open-shell restricted Hartree Fock Zombie state and the remaining 63 Zombie states generated using the same biasing method as before. Similarly, the first Zombie state can be set to the six electron restricted Hartree Fock Zombie state. It can be seen in \figr{imgbiased64HF67} that depending on the  choice of the initial condition the same ZS propagation can lead the ground state of neutral molecule and its anion. \figr{imgbiasedreduced} shows the result of the propagation using the basis of 10, 30 and 50 random basis ZS functions together with their improved energies obtained by cleaning. The first basis function in each basis set is the six electron restricted Hartree Fock determinant to ensure each evolution starts at the same energy. As expected the energies for these smaller biased bases are not as accurate as obtained using the 'complete' 64 basis function Hamiltonian. However, the biasing method is considerably more accurate when a directly comparing random and biased bases of the same size.

\begin{figure}[H]
\centering
 \includegraphics[width=\columnwidth]{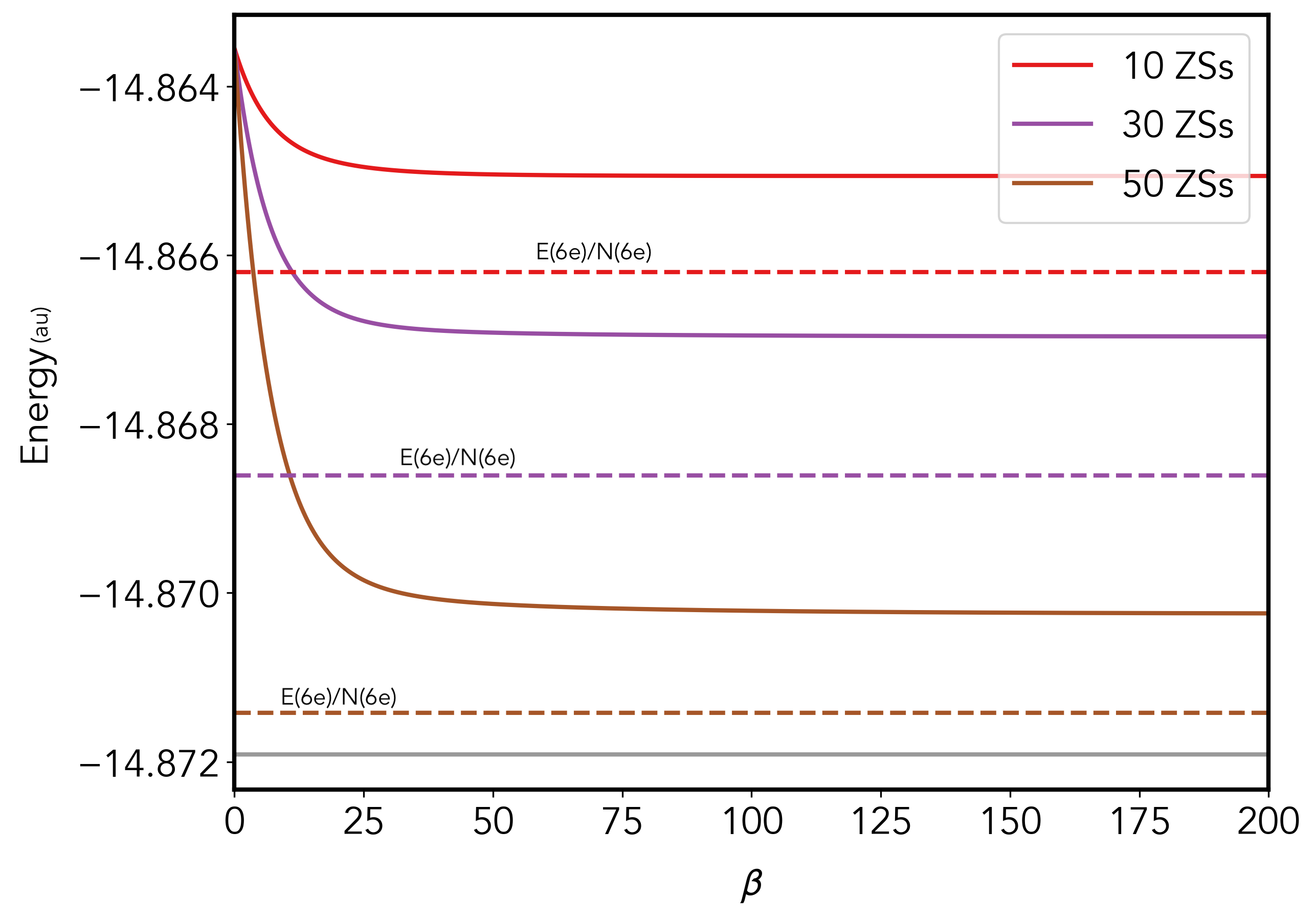}
 \caption{Imaginary time propagation for basis sets containing 50, 30 and 10 basis functions the first being the six electron restricted Hartree Fock determinant and the rest being biased Zombie state basis functions. The details of how each electron is biased are summarised in table \ref{table:biased}. The eigenvalue for the Li$_2$ ground state obtained by diagonalising the Slater determinant basis is given, as the solid horizontal line, for comparison. The cleaned energies, found by $E(6e)/N(6e)$, are shown as dashed lines.}
\figl{imgbiasedreduced}
\end{figure}

\begin{figure}[H]
    \centering
    \begin{subfigure}{\columnwidth}
    \centering
    \includegraphics{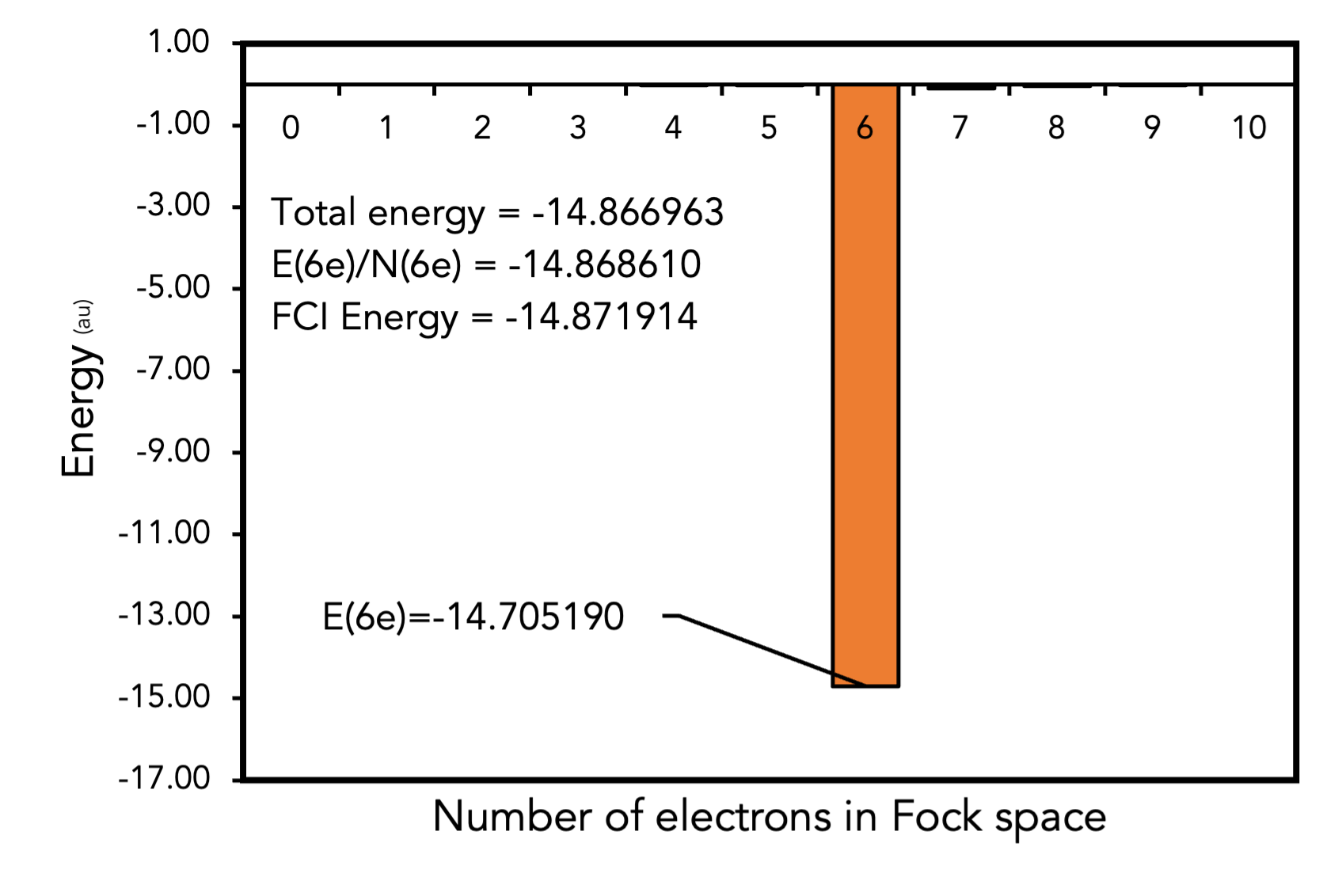}
    \end{subfigure}
    \begin{subfigure}{\columnwidth}
    \centering
    \includegraphics{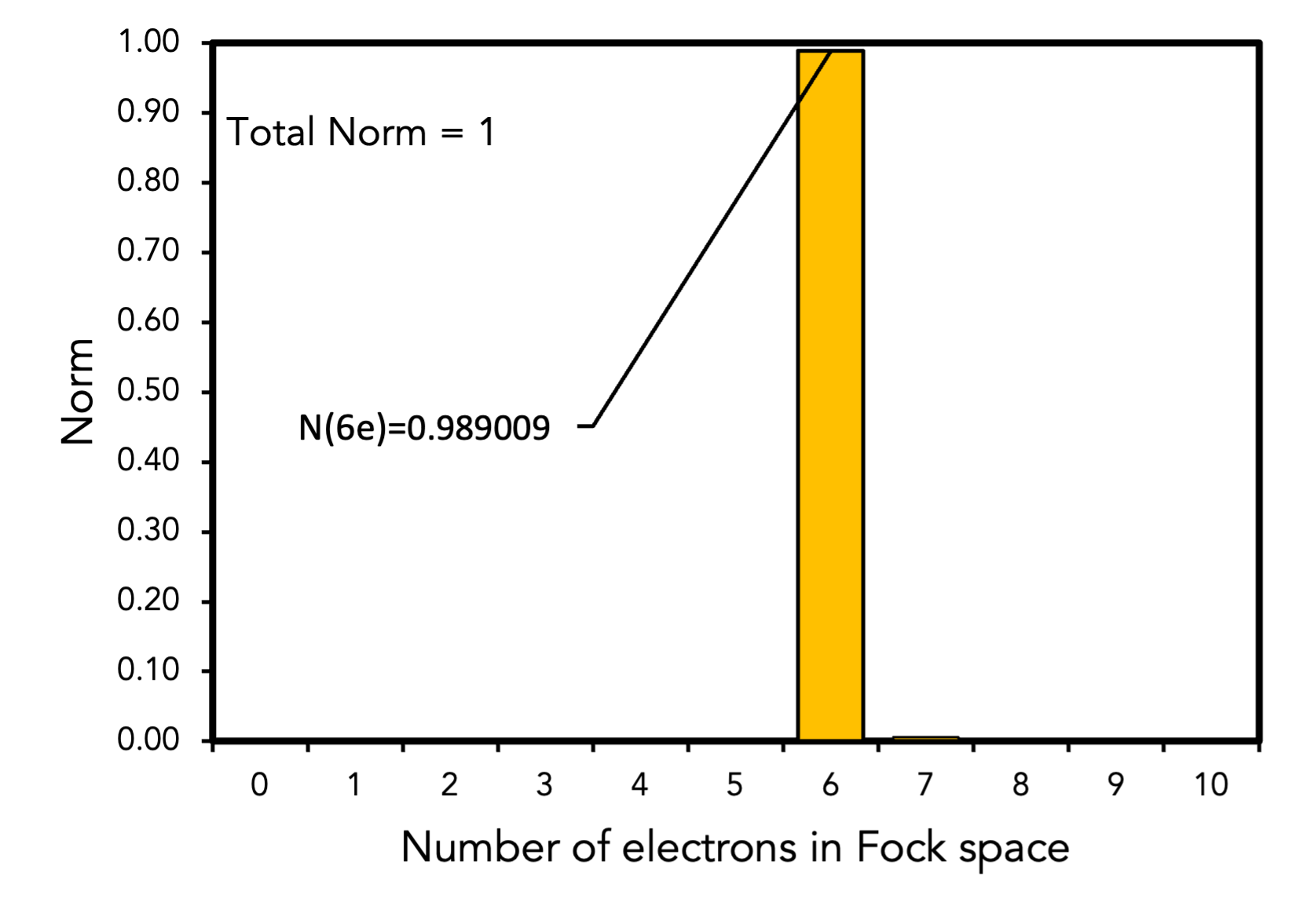}
    \end{subfigure}
    \caption{Energy (top) and norm (bottom) distribution for each number of electrons for a basis set of 30 biased Zombie states the first being set as the six electron restricted Hartree Fock determinant. The majority of the final wave function belongs to the Fock space on $n_e=6$ electrons.}
    \label{fig:biased30clean}
\end{figure}

\figr{biased30clean}  shows the distribution of energy and norm over the Fock spaces with $m_e$ electrons for the case of 30 BFs. Almost all norm is in the 6e Fock Space but even in this case cleaning improves the energy and brings it closer to the Full CI result.  Figure \figr{imgbiased64HF67} shows that a basis of 64 randomly selected basis functions is complete and yields the exact result even without cleaning.

\subsection{Low-lying excited states}
Gram Schmidt orthogonalisation was then applied to the complete random basis and is shown in \figr{gs_rand}. Four separate states were computed, all of which were initially set to arbitrary vectors. The Gram Schmidt process was applied and then each state was propagated in imaginary time with Gram Schmidt orthogonalization applied after every time step. It can be seen in \figr{gs_rand} that the first state is the neutral ground state the second and third states are the degenerate anion ground states (degenerate as $M_s$ can be $\pm 1/2$) and the fourth is the first neutral excited state for Li$_2$. All of these values match the corresponding eigenvalues found by diagonalising the complete Slater determinant basis.

The same Gram Schmidt process was then applied to four states using a 64 Zombie state biased Hamiltonian. The imaginary time propagation of these four states can be seen in \figr{gs_biased}. As with complete random basis propagating in imaginary time with Gram Schmidt produces the lowest neutral state, the degenerate anion ground states and the first excited energy which are also shown on both figures as dashed lines.
\begin{figure}[H]
\centering
 \includegraphics[width=\columnwidth]{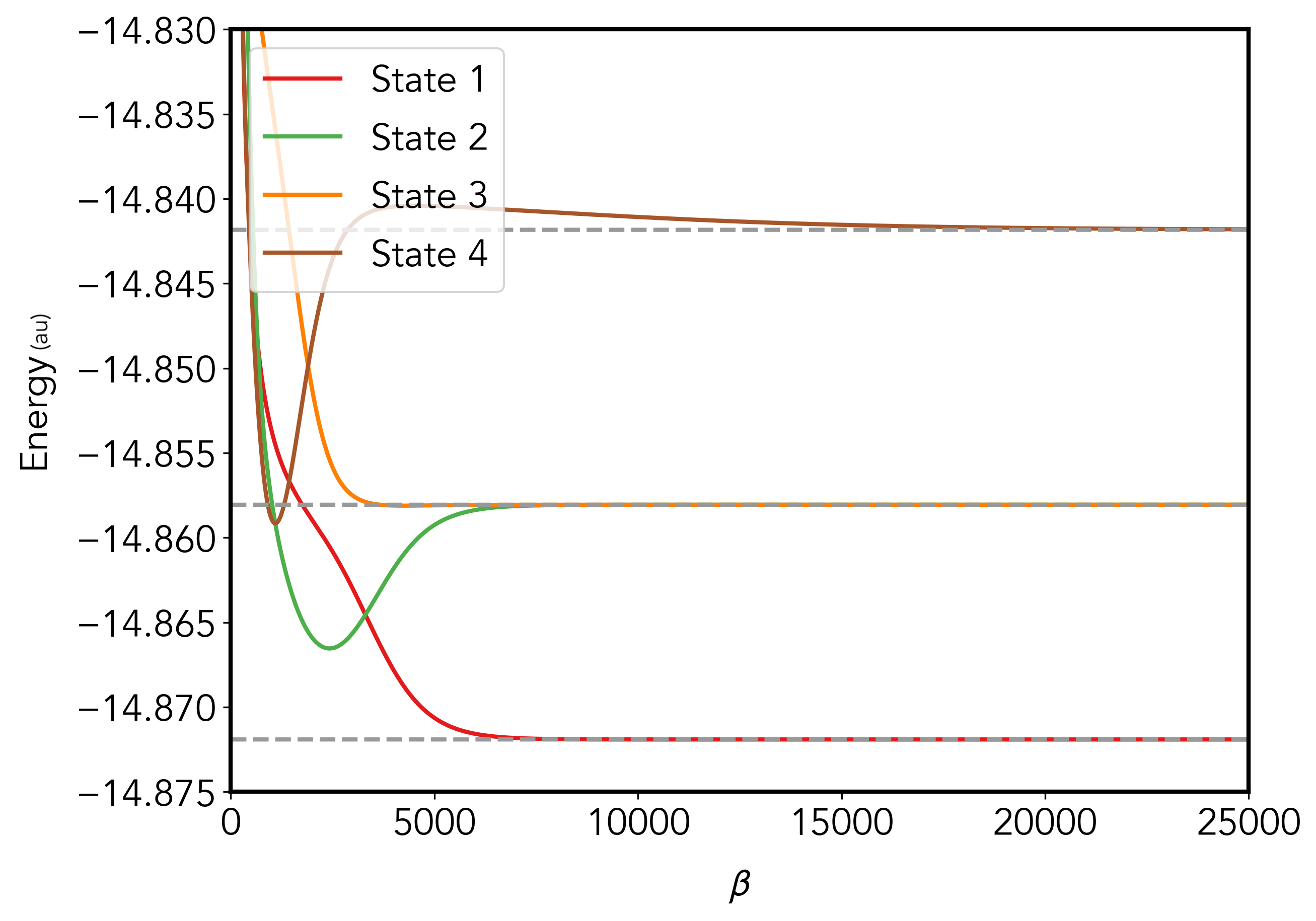}
\caption{Imaginary time propagation using Gram Schmidt orthogonalisation and the complete random basis for Li$_2$. The final energies of each state correspond accordingly: State 1 is the neutral ground state; states 2 and 3 are the degenerate anion states and state 4 is the first neutral excited state. These values equal the eigenvalues obtained by diagonalizing the complete Slater determinant basis which have been shown with the dashed lines.}
\figl{gs_rand}
\end{figure}

\begin{figure}[H]
\centering
 \includegraphics[width=\columnwidth]{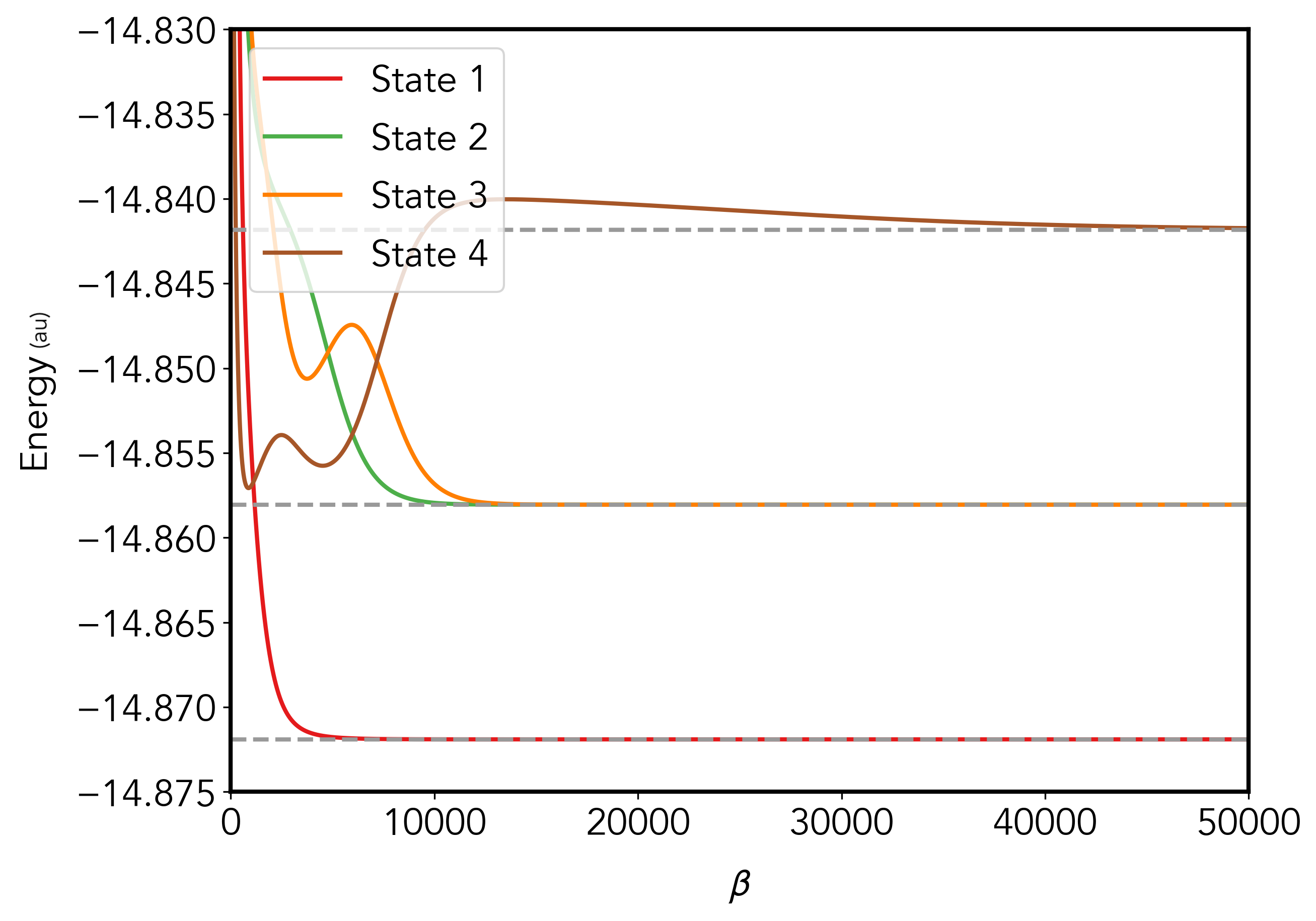}
 \caption{Imaginary time propagation using Gram Schmidt orthogonalisation and the 64 Zombie state biased basis functions for Li$_2$. The final energies of each state correspond accordingly: State 1 is the neutral ground state; states 2 and 3 are the anion ground state and state 4 is the first excited state. These values equal the eigenvalues obtained by diagonalising the complete Slater determinant basis which have been shown with the dashed lines.}
\figl{gs_biased}
\end{figure}

\section{Conclusions}

The findings of this paper are broad but substantially advance Zombie states as a potential method for simulating electronoic structure and dynamics. While previously shown to be numerically consistent, the alternative formalism of Zombie states, presented here, builds from the the vacuum state using creation and annihilation operators to create Zombie states. This forms a clearer link between Zombie states and more traditional electronic structure theory. This new formulation also leads to a stricter normalisation condition which not only demonstrated how ideas from bosonic systems can be translated to fermionic systems but was key in the implementation of the biasing method used. A key facet of a practicable modern methods are efficient algorithms for the calculation of the Hamiltonian and other operators. In this paper we have presented algorithms that greatly improve on the na\"ive algorithms for the number operator, spin operators ($\hat S_z$ and $\hat S^2$) and the two-electron Hamiltonian. A scaled algorithm for the two-electron Hamiltonian is presented and reduces computational cost significantly. Computationally inexpensive calculation of important system properties is obviously very important for any subsequent work. But this also sets up a easily replicable method for reducing the scaling of any algorithms that may be developed in subsequent work. All of the algorithms follow a similar method using the action of creation and annihilation operators to obtain recursion relations which lead to lower-scaling algorithms.

We have also shown that using imaginary time propagation on a system of Zombie states it is possible to find the lowest-lying state of that system. A complete basis of Slater determinants and a basis of complete random Zombie states were shown to give the exact same ground state energies for Li$_2$. Imaginary time propagation removes the need for long real-time evolution and Fourier transforms. Reducing the size of the random basis resulted in the ground state energy obtained no longer matching the benchmark set by the complete Slater determinant basis. However, this problem is overcome by the use of the biasing method, where we use the ideas of CASSCF\cite{casscf} (or CASCI) by splitting orbitals into inactive, active and virtual types and then biasing individual orbitals in the active space. The lowest four spin orbitals were set as core orbitals and so always "alive", leaving two electrons to fill the remaining six spin orbitals, such that there were $2^6=64$ Slater Determinants in a complete basis set for this active space and any more basis functions would lead to linear dependencies and a singularities in the overlap matrix. But with 64 biased Zombie state basis functions it was possible to produce, with imaginary time evolution, energy levels that matched the 16 times larger Slater determinant basis. We also showed that with a single Zombie state changed to a specified restricted Hartree Fock determinant it was possible to find a specific energy level with imaginary time propagation. Reducing the number of biased Zombie states gives less accurate energy levels, however, when compared to imaginary time propagation for random bases of the same size the biased bases are considerably better. In some respects the accuracy of calculations with 64 basis functions is not a great surprise from a chemistry viewpoint, since the lowest four electrons are in 1s orbitals, and therefore tightly bound to the Lithium nuclei. It would be energetically very costly to excite these electrons (compared to the 2s electrons), and they are therefore expected to remain core 1s in any low-lying states.  Clearly, the use of Zombie states in not giving new scientific insight into the electronic structure of Li$_2$ but we have demonstrated that ZS can reproduce results with a reduced basis in an efficient manner. 

It is worth noting that 64 basis functions is less than the 210 configurations possible with six electrons in ten spin orbitals but far more than the 15 possible configurations possible when the first four electrons are fixed and the remaining two can be arranged across six spin orbitals. Thus, introduction of Zombie States with non-integer occupations of orbitals for a small system has not decreased the required number of basis functions, but in fact increased them.  This is not surprising as the basis of Zombie States describes multi-electronic states with all possible numbers of electrons from no electrons at all to electrons occupying all spin-orbitals. For a systems with small active space,  like Li$_2$ for example,  the price of not sticking to the right number of electrons outweighs the benefit of bringing in other excited electronic configurations which may be disregarded by active space methods. However, as can be seen from the figure \figr{imgbiasedreduced} a small ZS basis accounts for some correlation energy. For larger systems, where the active space is large, this may be an affordable way of taking electron correlation into account.  Future work will focus on applying Zombie states to much larger systems that traditionally require extremely large basis sets, such as N$_2$,\cite{doi:10.1063/1.3302277} to obtain FCI values while using significantly smaller ZS basis sets. We will also aim to apply sampling techniques previously developed for Bosonic systems\cite{doi:10.1063/1.5020567} to Zombie states which should aid in converging results, and address the possibility of choosing a linear combination of zombie states such that the overall wavefunction is an eigenstate of the number operator.

\section*{Authors' Contributions}
Oliver A Bramley and Timothy J H Hele contributed equally to this work.

\section*{Author Declarations}
The authors have no conflicts to disclose.

\section*{Acknowledgements}
The authors would like to thank George Booth for discussions on PySCF and Dmitry Makhov for help with Molpro. The support of EPSRC via the Grant No. EP/P021123/1 is acknowledged. TJHH also acknowledges a Royal Society University Research Fellowship URF\textbackslash R1\textbackslash 201502.

\section*{Data Availability}
The data is available from corresponding author upon request

\appendix
\section{Number operator}

Although the observable properties of a zombie state can be formally computed exactly by iterative application of creation and annihilation operators followed by computation of the overlap of the resultant states\cite{RN106}, this usually is not an efficient calculation. In the appendices we present algorithms for commonly-used properties of electronic states such as the number of electrons, and their spin properties.

If we consider the action of the number operator 
\begin{align}
 \hat N = \sum_{m=1}^{M} \hat n_m =  \sum_{m=1}^{M} \hat b^{\dag}_m \hat b_m \eql{num}
\end{align}
where $\hat b_m$ is the annihilation operator and $\hat b^\dag_m$ is the creation operator for zombie state $m$, we see
\begin{align}
\hat n_m \ket{\bm{\zeta}^{(b)}} = 
\begin{bmatrix}
 a_{11}^{(b)} & a_{12}^{(b)} & \ldots & a_{1(m-1)}^{(b)} & a_{1m}^{(b)} & \ldots & a_{1M}^{(b)} \\
 a_{01}^{(b)} & a_{02}^{(b)} & \ldots & a_{0(m-1)}^{(b)} & 0 & \ldots & a_{0M}^{(b)}
\end{bmatrix}
\end{align}
such that $\hat n_m$ effectively `deletes' the coefficient $a_{0m}^{(b)}$. Note that the sign switching cancels itself out. It therefore appears that the acting with the number operator $\hat N$ is $\mathcal{O}(M)$ as the summation is over $M$ terms.

Comparison of \eqr{overlap} and \eqr{num} suggests that computing the number of electrons represented by any zombie state, i.e. $\bra{\bm{\zeta}^{(a)}} \hat N \ket{\bm{\zeta}^{(b)}}$ would be $\mathcal{O}(M^2)$ as computing the overlap of two Zombie states is $\mathcal{O}(M)$. 

However, it is possible to construct an algorithm which is only $\mathcal{O}(M)$ by careful consideration of the summation and the product, by adapting an algorithm from Ref. \onlinecite{hel11a}. Mathematically, calculation of the action of the number operator is
\begin{align}
 \bra{\bm{\zeta}^{(a)}} \hat N \ket{\bm{\zeta}^{(b)}} = & \sum_{l=1}^{M} \bra{\bm{\zeta}^{(a)}} \hat n_l \ket{\bm{\zeta}^{(b)}} \no \\
 = & \sum_{l=1}^{M} \left[ \prod_{j = 1}^{l-1} a_{1 j}^{(a)*}a_{1 j}^{(b)}+ a_{0 j}^{(a)*}a_{0 j}^{(b)} \right] a_{1l}^{(a)*}a_{1l}^{(b)} \left[ \prod_{j = l+1}^{M} a_{1 j}^{(a)*}a_{1 j}^{(b)}+ a_{0 j}^{(a)*}a_{0 j}^{(b)} \right]. \eql{numbig}
\end{align}

We can define
\begin{align}
    d_i=a^{(a)*}_{1i}a^{(b)}_{1i}
    \eql{numb_d}
\end{align}
and use \eqr{ham_low3} to define 
\begin{align}
    g_i = \prod_{i=1}^{m}f_i.
    \eql{numb_g}
\end{align}
This can be used along with \eqr{ham_low7} to find the following recursion relations:
\begin{subequations}
\begin{align}
g_1 = & f_1, \\
g_l = & g_{l-1}f_l,\quad  l = 2,3, \ldots M \\
h_M = & f_M, \\
h_m = & h_{l+1}f_l, \quad  l = M-1, M-2, \ldots, 1.
\end{align}
\end{subequations}

Computation of \eqr{numb_d} and \eqr{ham_low3} are clearly $\mathcal{O}(M)$ and using the recursion relations so are computation of $\{g_l\}$ and $\{h_m\}$.
Inserting the aforementioned equations into \eqr{numbig}
\begin{align}
 \bra{\bm{\zeta}^{(a)}} \hat N \ket{\bm{\zeta}^{(b)}} = \sum_{l=1}^{M} g_{l-1} d_l h_{l+1}
\end{align}
which is a summation that can also be computed in $\mathcal{O}(M)$ steps. When applied this new algorithm is significantly faster than the previous method; increasing the number of orbitals by a factor of 10, the new algorithm becomes roughly another 10 times faster than the old one as to be expected from the scaling arguments above. Full details of this code racing can be found in \textit{Table 5} of the supplementary material.

Since Zombie states are not usually eigenstates of the number operator, unlike most Slater determinants, the uncertainty in the in number of electrons can be defined as a standard deviation
\begin{align}
 \sigma_N = \sqrt{ \langle \hat N^2 \rangle - \langle \hat N\rangle }
\end{align}
For $\hat N^2$,
\begin{align}
 \bra{\zeta^{(a)}} \hat N^2 \ket{\zeta^{(b)}} = 
 & \sum_{k=1}^M \bra{\bm{\zeta}^{(a)}} \hat N  \hat  n_k \ket{\bm{\zeta}^{(b)}}
\end{align}
Computing $\hat n_k \ket{\bm{\zeta}^{(b)}}$ is trivial (simply delete the `dead' coefficient for orbital $k$). We can therefore define $\ket{\bm{\zeta}^{(b)}_k} = \hat n_k \ket{\bm{\zeta}^{(b)}}$ and compute
\begin{align}
  \bra{\bm{\zeta}^{(a)}} \hat N^2 \ket{\bm{\zeta}^{(b)}} = 
 & \sum_{k=1}^M \bra{\bm{\zeta}^{(a)}} \hat N  \ket{\bm{\zeta}^{(b)}_k}
\end{align}
using the known algorithm for computing the number operator, so this is $\mathcal{O}(M^2)$ instead of $\mathcal{O}(M^3)$.

The general idea presented trivially extends to the `Ghost' operator
\begin{align}
 \hat G = \sum_{m=1}^{M} \hat s_m :=  \sum_{m=1}^{M} \hat b_m \hat b^{\dag}_m \eql{dead}
\end{align}
which counts how many of the zombie states are `dead' (unoccupied), as opposed to the number operator which counts how many of them are alive.

\section{Spin operators}

\subsection{$\hat S_z$ operator}
Orbitals are defined with the assumption that any used in the zombie state calculation are from a \emph{restricted} calculation such that for each orbital with spin $\ket{\alpha}$ (spin up, $m_s = +1/2$) there exists an orbital with the same spatial wavefunction but spin component $\ket{\beta}$ (spin down, $m_s = -1/2$). For a system with $M$ spin orbitals, there will consequently be $K$ spatial orbitals where $K = M/2, K \in \mathbb{N}$. We further define that all spin orbitals with up spin have an odd index $m = 1,3,\ldots M-1$  and all spin orbitals with down spin have an even index $m = 2, 4,\ldots, M$, such that the $k$th spatial orbital has $\ket{\alpha}$ spin orbital $m=2k-1$ and $\ket{\beta}$ spin orbital $m=2k$, $k = 1, 2, \ldots, K$.

Using this numbering convention, in second quantization notation
\begin{subequations}
\begin{align}
 \hat S_z = & \frac{1}{2}\sum_{k=1}^{K} \hat n_{2k-1} - \hat n_{2k} \\
 = & \frac{1}{2} \sum_{m=1}^{M} (-1)^{m-1}\hat n_m.
\end{align}
\eql{szeq}
\end{subequations}

The optimised number operator is then adapted introducing a simple sign change rule to calculate $S_z$ in $\mathcal{O}(M)$ steps. Comparison of the na\"ive algorithm, using creation and annihilation operators, and the optimised algorithm, co-opting the optimised number operator equations, can be found in \textit{Table 5} of the supplementary material.

\subsection{Faster $\hat S_z^2$ computation}
Prima facie, computation of $\hat S_z^2$ would be $\mathcal{O}(M^3)$ using \eqr{szeq}
\begin{align}
 \hat S_z^2 = & \frac{1}{4}\sum_{k}^{M/2} \sum_{j}^{M/2}(\hat n_{2k-1} - \hat n_{2k})(\hat n_{2j-1} - \hat n_{2j}) = \frac{1}{4} \sum_{k}^{M} \sum_{j}^{M} \hat n_k \hat n_j (-1)^{j+k}
\end{align}
so there is $\mathcal{O}(M)$ for the summation over $k$, another over $j$, and another to evaluate overlap.

However, extending the adapted number operator algorithm used for $\hat S_z$ in $\mco(M)$  calculation to $\hat S_z^2$ is straightforward, 
\begin{align}
 \bra{\bm{\zeta}^{(a)}} \hat S_z^2 \ket{\bm{\zeta}^{(b)}} = \frac{1}{2} \sum_{m}^M \bra{\bm{\zeta}^{(a)}} \hat S_z \hat n_m \ket{\bm{\zeta}^{(b)}} (-1)^m
\end{align}
but $\hat n_m \ket{\bm{\zeta}^{(b)}}$ gives another zombie state that we call $\ket{\bm{\zeta}^{(b,m)}}$ such that 
\begin{align}
 \bra{\bm{\zeta}^{(a)}} \hat S_z^2 \ket{\bm{\zeta}^{(b)}} = \frac{1}{2} \sum_{m}^M \bra{\bm{\zeta}^{(a)}} \hat S_z \ket{\bm{\zeta}^{(b,m)}} (-1)^m.
\end{align}
The bra-ket can be evaluated in $\mco(M)$ and therefore the action of the $\hat S_z^2$ operator in $\mco(M^2)$.

\subsection{Total spin}
Usually in electronic structure calculation one also wants to know the total spin\cite{RN111}
\begin{subequations}
\begin{align}
 \hat S^2 = &  \hat S_x^2 + \hat S_y^2 + \hat S_z^2 \\
 = & \hat S_+ \hat S_- - \hat S_z + \hat S_z^2
\end{align}
\end{subequations}

We can apply the faster $\hat S_z$ and $\hat S_z^2$ algorithms as detailed above. $\hat S_+$ and $\hat S_-$ are raising and lowering operators,
\begin{subequations}
\begin{align}
 \hat S_+ = & \sum_{k=1}^{N_{\rm spa}} \hat s_{+,k} \\
 \hat S_- = & \sum_{k=1}^{N_{\rm spa}} \hat s_{-,k} \\
 \hat s_{+,k} = & \hat b_{2k-1}^\dag \hat b_{2k} \\
 \hat s_{-,k} = & \hat b_{2k}^\dag \hat b_{2k-1}
\end{align}
\end{subequations}
where $K$ is the number of \emph{spatial obitals}, viz $K = M/2$. We are numbering the orbtials $1, 2, \ldots, K$, such that the $k$th spatial orbital has an alpha spin orbital number $2k-1$ and a beta spin orbital number $2k$.

We then observe the effect of $\hat s_{+,k}$ and $\hat s_{-,k}$ on the wavefunction (where $m=2k$)
\begin{subequations}
\begin{align}
\hat s_{+,k} \ket{\bm{\zeta}^{(b)}} =  \hat b^\dag_{2k-1} \hat b_{2k} &
\begin{bmatrix}
 a_{11}^{(b)} & a_{12}^{(b)} & \ldots & a_{1(m-1)}^{(b)} & a_{1m}^{(b)} & a_{1(m+1)}^{(b)} & \ldots & a_{1M}^{(b)} \\
 a_{01}^{(b)} & a_{02}^{(b)} & \ldots & a_{0(m-1)}^{(b)} & a_{0m}^{(b)} & a_{0(m+1)}^{(b)} & \ldots & a_{0M}^{(b)} 
\end{bmatrix} \\
=  \hat b^\dag_{2k-1}  &
\begin{bmatrix}
 -a_{11}^{(b)} & -a_{12}^{(b)} & \ldots & -a_{1(m-1)}^{(b)} & 0 & a_{1(m+1)}^{(b)} & \ldots & a_{1M}^{(b)} \\
 a_{01}^{(b)} & a_{02}^{(b)} & \ldots & a_{0(m-1)}^{(b)} & a_{1m}^{(b)} & a_{0(m+1)}^{(b)} & \ldots & a_{0M}^{(b)} 
\end{bmatrix}
\\
= & \begin{bmatrix}
 a_{11}^{(b)} & a_{12}^{(b)} & \ldots & a_{0(m-1)}^{(b)} & 0 & a_{1(m+1)}^{(b)} & \ldots & a_{1M}^{(b)} \\
 a_{01}^{(b)} & a_{02}^{(b)} & \ldots & 0 & a_{1m}^{(b)} & a_{0(m+1)}^{(b)} & \ldots & a_{0M}^{(b)} 
\end{bmatrix}
\end{align}
\end{subequations}
and
\begin{subequations}
\begin{align}
 \hat s_{-,k} \ket{\bm{\zeta}^{(b)}} =  \hat b^\dag_{2k} \hat b_{2k-1} &
\begin{bmatrix}
 a_{11}^{(b)} & a_{12}^{(b)} & \ldots & a_{1(m-1)}^{(b)} & a_{1m}^{(b)} & a_{1(m+1)}^{(b)} & \ldots & a_{1M}^{(b)}\\
 a_{01}^{(b)} & a_{02}^{(b)} & \ldots & a_{0(m-1)}^{(b)} & a_{0m}^{(b)} & a_{0(m+1)}^{(b)} & \ldots & a_{0M}^{(b)} 
\end{bmatrix}
\\
=  \hat b^\dag_{2k} & 
\begin{bmatrix}
 -a_{11}^{(b)} & -a_{12}^{(b)} & \ldots & 0 & a_{1m}^{(b)} & a_{1(m+1)}^{(b)} & \ldots & a_{1M}^{(b)} \\
 a_{01}^{(b)} & a_{02}^{(b)} & \ldots & a_{1(m-1)}^{(b)} & a_{0m}^{(b)} & a_{0(m+1)}^{(b)} & \ldots & a_{0M}^{(b)} 
\end{bmatrix}
\\
= & \begin{bmatrix}
a_{11}^{(b)} & a_{12}^{(b)} & \ldots & 0 & a_{0m}^{(b)} & a_{1(m+1)}^{(b)} & \ldots & a_{1M}^{(b)} \\
 a_{01}^{(b)} & a_{02}^{(b)} & \ldots & a_{1(m-1)}^{(b)} & 0 & a_{0(m+1)}^{(b)} & \ldots & a_{0M}^{(b)} 
\end{bmatrix}
\end{align} \eql{loweringop}
\end{subequations}
In both cases the sign changes cancel each other out exactly. This means that no sign change is required for evaluating $\hat S_+$ or $\hat S_-$, saving computational cost. However, there are clearly reductions to be made in the scaling that should reduce the time needed to compute $\bra{\bm{\zeta}^{(a)}} \hat s_{+,l} \hat s_{-,k} \ket{\bm{\zeta}^{(b)}}$. The action of $\hat s_{-,k} \ket{\bm{\zeta}^{(b)}}$ defined by \eqr{loweringop} can be defined as $\hat s_{-,k} \ket{\bm{\zeta}^{(b)}}=\ket{\bm{\zeta}_k^{(c)}}$ (where $ K = M/2$ and $m=2l$) which has $\mco(M)$ scaling. And so it is possible to compute

\begin{equation}
\begin{aligned}
 &\bra{\bm{\zeta}^{(a)}} \hat S_{+} \hat S_{-} \ket{\bm{\zeta}^{(b)}} = \sum_{k=1}^{K} \bra{\bm{\zeta}^{(a)}} \hat S_{+} \ket{\bm{\zeta}^{(c)}_k}= \sum_{k=1}^{K} \sum_{l=1}^{K}\bra{\bm{\zeta}^{(a)}} \hat s_{+,m} \ket{\bm{\zeta}^{(c)}_k}\\
 &= \sum_{k=1}^{K} \left[ \sum_{l=1}^{K} \left[\prod_{j=1}^{l-2} a_{1j}^{(a)*}a_{1j}^{(c)} + a_{0j}^{(a)*}a_{0j}^{(c)} \right]\cdot a_{0(l-1)}^{(a)*}a_{1(l-1)}^{(c)} \cdot a_{1l}^{(a)*}a_{0l}^{(c)} \cdot \left[\prod_{j=l+1}^{M} a_{1j}^{(a)*}a_{1j}^{(c)} + a_{0j}^{(a)*}a_{0j}^{(c)} \right] \right]
\end{aligned} 
\end{equation}

In a similar manner to the scaled Hamiltonian with, \eqr{ham_low3} and the number operator with \eqr{numb_g} and \eqr{numb_d} we define
\begin{subequations}
\begin{align}
 f_i = & (a_{1 i-1}^{(a)*}a_{1 i-1}^{(c)} + a_{0 i-1}^{(a)*}a_{0 i-1}^{(c)})\cdot(a_{1 i}^{(a)*}a_{1 i}^{(c)} + a_{0 i}^{(a)*}a_{0 i}^{(c)})\\
 g_l = & \prod_{i=1}^{l} f_i \\
 h_l = & \prod_{i=l}^{K} f_i \\
 d_i = & a_{0(i-1)}^{(a)*}a_{1(i-1)}^{(c)}a_{1i}^{(a)*}a_{0i}^{(c)}
\end{align}
\end{subequations}

Which gives
\begin{align}
\bra{\bm{\zeta}^{(a)}} \hat s_{+l} \ket{\bm{\zeta}^{(c)}_k}=g_{l-2}d_{l}h_{l+1}
\end{align}

Overall this reduces the scaling from  $\mco(M^3)$ to  $\mco(M^2)$. This new algorithm was then used alongside the original speed improvements giving a large time speedup, see \textit{Table 8} in the supplementary material for full details. However, it should be possible to scale the $\bra{\bm{\zeta}^{(a)}} \hat s_{+} \hat s_{-} \ket{\bm{\zeta}^{(b)}}$ removing the need to calculate $\ket{\bm{\zeta}^{(c)}}$. We can set $m=2l$ and $n=2k$ and then define 
\begin{align}
\hat s_{+,l} \hat s_{-,k} \ket{\bm{\zeta}^{(b)}} =  \hat b^\dag_{2l-1} \hat b_{2l} \hat b^\dag_{2k} \hat b_{2k-1} 
\begin{bmatrix}
 a_{11}^{(b)} & a_{12}^{(b)} & \ldots & a_{1m}^{(b)} & \ldots & a_{1n}^{(b)}& \ldots & a_{1M}^{(b)} \\
 a_{01}^{(b)} & a_{02}^{(b)} & \ldots & a_{0m}^{(b)} & \ldots & a_{0n}^{(b)}& \ldots & a_{0M}^{(b)} 
\end{bmatrix} 
\end{align}

which can have the following outcomes
\begin{equation}
\hat s_{+,l} \hat s_{-,k} \ket{\bm{\zeta}^{(b)}} = \begin{cases}
\begin{bmatrix}
 a_{11}^{(b)} & a_{12}^{(b)}  & \ldots & 0 &  a_{0n}^{(b)} & \ldots & a_{0(m-1)}^{(b)} & 0 & \ldots & a_{1M}^{(b)} \\
 a_{01}^{(b)} & a_{02}^{(b)} &  \ldots & a_{1(n-1)}^{(b)}& 0 & \ldots & 0 & a_{1m}^{(b)}  & \ldots & a_{0M}^{(b)}
\end{bmatrix} & \text{if $2k<2l$;}\\[5ex]
\begin{bmatrix}
 a_{11}^{(b)} & a_{12}^{(b)} & \ldots & a_{0(m-1)}^{(b)} & 0  & \ldots & 0 &  a_{0n}^{(b)} & \ldots & a_{1M}^{(b)} \\
 a_{01}^{(b)} & a_{02}^{(b)} & \ldots & 0 & a_{1m}^{(b)}  & \ldots & a_{1(n-1)}^{(b)}& 0  & \ldots & a_{0M}^{(b)}
\end{bmatrix} & \text{if $2k>2l$;}\\[5ex]
\begin{bmatrix}
a_{11}^{(b)} & a_{12}^{(b)}&\ldots&a_{1(m-2)}^{(b)}&a_{1(m-1)}&0& a_{1(m+1)}^{(b)} & \ldots & a_{1M}^{(b)} \\
a_{01}^{(b)} & a_{02}^{(b)}&\ldots&a_{0(m-2)}^{(b)}&0&a_{0m}& a_{0(m+1)}^{(b)} & \ldots & a_{0M}^{(b)}
\end{bmatrix}& \text{if $2k=2l$}
\end{cases}
\end{equation}

which gives
\begin{subequations}
\begin{align}
\bra{\bm{\zeta}^{(a)}}\hat s_{+,l} \hat s_{-,k}\ket{\bm{\zeta}^{(b)}} =\begin{cases}
\left[ \prod_{j = 1}^{k-2} a_{1 j}^{(a)*}a_{1 j}^{(b)}+ a_{0 j}^{(a)*}a_{0 j}^{(b)}\right]a_{0(k-1)}^{(a)*}a_{1(k-1)}^{(b)}a_{1k}^{(a)*}a_{0k}^{(b)}\\
\cdot\left[ \prod_{j = k+1}^{l-2} a_{1 j}^{(a)*}a_{1 j}^{(b)}+ a_{0 j}^{(a)*}a_{0 j}^{(b)}\right]a_{1(l-1)}^{(a)*}a_{0(l-1)}^{(b)}a_{0l}^{(a)*}a_{1l}^{(b)}\\ 
\cdot\left[ \prod_{j=l+1}^{M} a_{1 j}^{(a)*}a_{1 j}^{(b)}+ a_{0 j}^{(a)*}a_{0 j}^{(b)}\right] & \text{if $2k<2l$;}\\[3ex]
\end{cases}\\
\bra{\bm{\zeta}^{(a)}}\hat s_{+,l} \hat s_{-,k}\ket{\bm{\zeta}^{(b)}} =\begin{cases}
\left[ \prod_{j = 1}^{l-2} a_{1 j}^{(a)*}a_{1 j}^{(b)}+ a_{0 j}^{(a)*}a_{0 j}^{(b)}\right]a_{1(l-1)}^{(a)*}a_{0(l-1)}^{(b)}a_{0l}^{(a)*}a_{1l}^{(b)}\\
\cdot\left[ \prod_{j=l+1}^{k-2} a_{1 j}^{(a)*}a_{1 j}^{(b)}+ a_{0 j}^{(a)*}a_{0 j}^{(b)}\right]a_{0(k-1)}^{(a)*}a_{1(k-1)}^{(b)}a_{1k}^{(a)*}a_{0k}^{(b)}\\ 
\cdot\left[ \prod_{j=k+1}^{M} a_{1 j}^{(a)*}a_{1 j}^{(b)}+ a_{0 j}^{(a)*}a_{0 j}^{(b)}\right]& \text{if $2k>2l$;}\\[3ex]
\end{cases}\\
\bra{\bm{\zeta}^{(a)}}\hat s_{+,l} \hat s_{-,k}\ket{\bm{\zeta}^{(b)}} =\begin{cases}
\left[\prod_{j = 1}^{l-2} a_{1 j}^{(a)*}a_{1 j}^{(b)}+ a_{0 j}^{(a)*}a_{0 j}^{(b)}\right]a_{1(l-1)}^{(a)*}a_{1(l-1)}^{(b)}a_{0l}^{(a)*}a_{0l}^{(b)}\\
\cdot\left[ \prod_{j = l+1}^{M} a_{1 j}^{(a)*}a_{1 j}^{(b)}+ a_{0 j}^{(a)*}a_{0 j}^{(b)}\right] &\text{if $2k=2l$}
\end{cases}
\end{align} 
\end{subequations}

We then have to define the following recursion relations
\begin{subequations}
\begin{align}
 f_i = & (a_{1(i-1)}^{(a)*}a_{1(i-1)}^{(b)}+a_{0(i-1)}^{(a)*}a_{0(i-1)}^{(b)})\cdot(a_{1(i)}^{(a)*}a_{1(i)}^{(b)}+a_{0(i)}^{(a)*}a_{0(i)}^{(b)})\\
 c_i = & a_{0(i-1)}^{(a)*}a_{1(i-1)}^{(b)}\cdot a_{1 i}^{(a)*}a_{0 i}^{(b)} \\
 d_i = & a_{1(i-1)}^{(a)*}a_{0(i-1)}^{(b)} \cdot a_{0 i}^{(a)*}a_{1 i}^{(b)}\\
 s_i = & a_{1(i-1)}^{(a)*}a_{1(i-1)}^{(b)}a_{0i}^{(a)*}a_{0i}^{(b)}.
\end{align}
\end{subequations}

These can all trivially be calculated in $\mcob{M}$ steps. We then also define
\begin{subequations}
\begin{align}
g_l= & \prod_{i=1}^{l} f_i\eql{tot_spina}\\ 
h_l= & \prod_{i=l}^{K} f_i\eql{tot_spinb}\\ 
t_{(l,p)} = & \prod_{i=l}^{p} f_i \eql{tot_spinc}
\end{align}
\end{subequations}

\eqr{tot_spina} and \eqr{tot_spinb} can can be calculated recursively in $\mcob{M}$ steps and \eqr{tot_spinc} in $\mcob{M^2}$. Therefore
\begin{equation}
\begin{aligned}
\bra{\bm{\zeta}^{(a)}}\hat S_{+} \hat S_{-} \ket{\bm{\zeta}^{(b)}}= &\sum_{k=1}^{K-1} \sum_{l=k+1}^{K} g_{k-1}c_kt_{(k+1,l-1)}d_lh_{l+1} + \\
&\sum_{l=1}^{K-1} \sum_{k=l+1}^{K} g_{l-1}d_lt_{(l+1,k-1)}c_kh_{k+1} + \sum_{l=1}^{K} g_{l-1}s_lh_{l+1}
\end{aligned}
\end{equation}

The na\"ive total spin algorithm was made up of three separate algorithms $\hat S_+ \hat S_-$, $\hat S_z$, $\hat S_z^2$ that scaled $\mcob{M^3}$, $\mcob{M^2}$ and $\mcob{M^3}$ respectively. The scaled algorithm reduced the scaling of each algorithm to $\mcob{M^2},$ $\mcob{M}$ and $\mcob{M^2}$ respectively. This scaling improvement resulted in $\hat S^2$ for 1000 orbitals being calculated over 460 times faster than the na\"ive algorithm; full code racing details are given in \textit{Table 8} and \textit{Table 9} of the supplementary material.

\bibliography{ref}

\end{document}


\maketitle
\bibliographystyle{aip}

\section{Algorithm Speed Testing}

\subsection{Two electron Hamiltonian}
Although Li$_2$ has been used throughout this paper N$_2$ using the $6-31$G$^{**}$ basis has been used for testing the two electron Hamiltonian matrix element algorithms. This is so larger numbers of orbitals could be tested while not having to calculate more two-electron integrals.

\begin{table}[H]
\centering
\begin{tabular}{c|c|c|c}
Number of orbitals $M$  & Nai\"ve    & Ignore zeros      & Speed up \\ \hline
10        & 0.246568 & 0.06343558  & 3.89     \\
50        & 463.9408 & 29.23557546 & 15.87   
\end{tabular}
\caption{Time (in seconds) to compute a two-electron Hamiltonian matrix element between two Zombie states with either 10 or 50 orbitals for N$_2$ in the $6-31$G$^{**}$ basis. The na\"ive algorithm loops through all indices the second algorithm only applies creation and annihilation operators when $\bk{ij}{kl} \neq 0$. This gives a good speed improvement.}
\end{table}

\begin{table}[H]
\centering
\begin{tabular}{c|c|c|c}
Number of orbitals $M$ & Nai\"ve    & Ignore zeros and better loops      & Speed up \\ \hline
10        & 0.246568 & 0.06304387  & 3.91     \\
50        & 463.9408 & 27.93265575 & 16.61   
\end{tabular}
\caption{Time (in seconds) to compute a two-electron Hamiltonian matrix element between two Zombie states with either 10 or 50 orbitals for N$_2$ in the $6-31$G$^{**}$ basis. The na\"ive algorithm loops through all indices the second algorithm only loops over combinations of $ijkl$ already known to be nonzero and applies the previous $\bk{ij}{kl} \neq 0$ rule before applying the creation and annihilation operators. This gives a good improvement on the na\"ive algorithm and slight improvement on the algorithm above that does not have improved looping.}
\end{table}

\begin{table}[H]
\centering
\begin{tabular}{c|c|c|c}
Number of orbitals $M$& Nai\"ve    & Precomputation       & Speed up \\ \hline
10        & 0.246568 & 0.06676888  & 3.69     \\
50        & 463.9408 & 49.16095242 & 9.44   
\end{tabular}
\caption{Time (in seconds) to compute a two-electron Hamiltonian matrix element between two Zombie states with either 10 or 50 orbitals for N$_2$ in the $6-31$G$^{**}$ basis. The na\"ive algorithm loops through all indices the second algorithm precomputes $\{ \ket{\zeta^{(a)}_{ij}} \}$ and $\{ \ket{\zeta^{(b)}_{lk}} \}$ and applies a rule that if any $a_{0 m}^{(a)*}a_{0 m}^{(b)} + a_{1 m}^{(a)*}a_{1 m}^{(b)}$ terms are zero, then $\Omega_{ab} = 0$. This gives a good improvement on the na\"ive algorithm for 10 orbitals but the time improvement is not as good for 50 orbitals as the previous algorithms.}
\end{table}

\begin{table}[H]
\centering
\begin{tabular}{c|c|c|c}
Number of orbitals $M$ & Nai\"ve    & Scaled      & Speed up \\ \hline
10        & 0.246568 & 0.01157104  & 21.31     \\
50        & 463.9408 & 4.50038746 & 103.09   
\end{tabular}
\caption{Time (in seconds) to compute a two-electron Hamiltonian matrix element between two Zombie states with either 10 or 50 orbitals for N$_2$ in the $6-31$G$^{**}$ basis. The na\"ive algorithm loops through all indices and has  $\mathcal{O}(M^5)$ the scaled algorithm has $\mathcal{O}(M^4)$ and offers the greatest speedup of all two-electron Hamiltonian algorithms.}
\end{table}

\subsection{Number operator}
\begin{table}[H]
\centering
\begin{tabular}{c|c|c|c}
Number of orbitals $M$ & Old algorithm & New Algorithm & Speedup \\ \hline
100 &  0.028711 & 0.00056004 & 51.27 \\
1000 & 3.0294 & 0.0051531 & 587.87
\end{tabular}
\caption{Time (in seconds) to compute orbital occupancy of random zombie state populated with gaussian deviates. The old algorithm is much slower in absolute terms, and shows worse scaling.}
\label{tab:speedupnum}
\end{table}

\subsection{Spin operators}
\begin{table}[H]
\centering
\begin{tabular}{c|c|c|c}
Number of orbitals $M$ & Slow $\hat S_z$& Fast $\hat S_z$ & Speedup \\ \hline
100 & 0.029760  & 0.00062376 &47.71 \\
1000 & 3.7123 &0.0063864  &581.28 
\end{tabular}
\caption{Time (in seconds) to compute $S_z$ of random zombie state populated with Gaussian deviates using the unscaled and faster scaled $\hat S_z$ algorithm. The new algorithm clearly demonstrates better scaling and is faster in absolute terms.}
\label{tab:sz}
\end{table}

\begin{table}[H]
\centering
\begin{tabular}{c|c|c|c}
Number of orbitals $M$ & Slow $\hat S_z^2$& Fast $\hat S_z^2$ & Speedup \\ \hline
100 & 6.6990  & 0.23865 &28.07 \\
1000 & 3554.8 &6.8003  &522.74 
\end{tabular}
\caption{Time (in seconds) to compute $S_z^2$ of random zombie state populated with Gaussian deviates using the unscaled and faster scaled $\hat S_z^2$ algorithm. The new algorithm clearly demonstrates better scaling and is faster in absolute terms.}
\label{tab:sz2}
\end{table}

\begin{table}[H]
\centering
\begin{tabular}{c|c|c|c|c|c}
\multirow{2}{*}{Number of orbitals} & \multirow{2}{*}{Na\"ive} & \multirow{2}{*}{Scale reduction} & \multirow{2}{*}{Prefactor reduction} & \multicolumn{2}{c}{Speedups Na\"ive with} \\
                                    &                        &                                  &                                      & Scale reduction           & Prefactor reduction           \\ \hline
100                                 & 0.85121                & 0.055180                         & 0.025146                             & 15.43                     & 33.85                         \\
1000                                & 792.20                 & 5.2828                           & 2.3451                               & 149.96                    & 337.81                       
\end{tabular}

\caption{Time (in seconds) to compute $\bra{\zeta^{(a)}}\hat S_{+} \hat S_{-} \ket{\zeta^{(b)}}$ of random zombie state populated with gaussian deviates. The nai\"ve algorithm uses the no sign change rule for $\hat S_+$ or $\hat S_-$ the scale reduced algorithm requires calculation of $\ket{\zeta^{(c)}}$ and the prefactor reduced algorithm directly computes $\hat S_+$ and $\hat S_-$ acting on $\ket{\zeta^{(b)}}$.  The speedups are for each scaled algorithm when compared to the nai\"ve algorithm orbitlas. Both scaled algorithms show a real time improvement with the prefactor reduced algorithm being twice as fast as the scaled algorithm while still having the same scaling.}
\label{tab:speeduptsp2}
\end{table}

\begin{table}[H]
\centering
\begin{tabular}{c|c|c|c}
Number of orbitals $M$ & Original & New    & Speedup \\
\hline
100                & 6.4861   &0.092060  & 70.45   \\
1000               & 4106.4   &8.77818    &467.60        
\end{tabular}
\caption{Time (in seconds) to compute total spin of random zombie state. The original algorithm uses none of the scaled algorithms whereas the new algorithm uses the scaled $\hat S_z^2$ and the fastest algorithm to calculate $\bra{\zeta^{(a)}}\hat S_{+} \hat S_{-} \ket{\zeta^{(b)}}$. The new algorithm is considerably faster than the original implementation and shows much better scaling.}
\label{tab:speedupTS3}    
\end{table}

\section{Generating Li$_2$ one and two electron integrals}
The one and two electron integrals were generated using PySCF\cite{PYSCF} with the following python code

\begin{lstlisting}[language=Python]
from pyscf import gto, scf

mol = gto.M(
    unit = 'Bohr',
    atom = 'Li 0 0 0; Li 0 0 6', 
    basis = '6-31g**',
    verbose = 1,
    symmetry = True,
    symmetry_subgroup = 0, 
)
myhf = scf.RHF(mol)

\end{lstlisting}

\bibliography{ref}